\documentclass[aps,prx,preprint,groupedaddress]{revtex4-1}

\usepackage{amsmath}
\numberwithin{equation}{section}
\usepackage{amsmath}
\usepackage{amssymb}
\usepackage{stmaryrd}
\usepackage{color}

\usepackage{graphicx}
\usepackage{bm}
\usepackage{braket}
\begin{document}
	\title{Theory of giant diode effect in 
		$d$-wave superconductor junctions on the surface of topological insulator}
	\author{Yukio Tanaka$^{1}$, Bo Lu$^{2}$, and Naoto Nagaosa$^{3,4}$}
	\affiliation{$^1$Department of Applied Physics, Nagoya University, Nagoya, 464-8603,Japan \\ 
		$^2$ Center for Joint Quantum Studies and Department of Physics, Tianjin University, Tianjin 300072, China \\
		$^3$ Center for Emergent Matter Science (CEMS), RIKEN, Wako, Saitama 351-0198, Japan \\ 
		$^4$ Department of Applied Physics, The University of Tokyo, Tokyo 113-8656, Japan }
	\date{\today}
	\begin{abstract}
Nonreciprocal responses of noncentrosymmetric quantum materials  		attract recent intensive interests, which is essential for the 
rectification function in diodes. A recent breakthrough is the 
discovery of superconducting diode effect. The principle to 
enlarge rectification effect is highly desired to guide the 
design of superconducting diode. Here, we study theoretically the Josephson 
junction S/FI/S (S: $d$-wave superconductor, FI: ferromagnetic 
insulator) on the surface of a topological insulator (TI). 
The simultaneous existence of $\sin\varphi$, $\cos\varphi$ and $\sin 2\varphi$ 	terms with almost the same order in Josephson current $I(\varphi)$ 
is essential to get larger values of $Q$ factor given by 
$Q=(I_{c}^{+} - \mid I_{c}^{-} \mid)/ (I_{c}^{+} + \mid I_{c}^{-} \mid)$
with $I_{c}^{+}={\rm max}(I(\varphi))$ and the negative one $I_{c}^{-}$
for macroscopic phase difference $\varphi$ of two superconductors on TI. 
We find that it can show a very large diode effect by tuning the crystal axes of $d$-wave superconductors and the magnetization of FI. The difference of 
the maximum Josephson currents $I_{c}$'s between the positive and negative
directions can be about factor 2, where the current-phase relation is 
modified largely from the conventional one. 
The relevance of the 
zero energy Andreev bound states 
as Majorana bound states
 at the interface is also revealed. This result can pave a way to realize an efficient superconducting diode with low energy cost.
\end{abstract}
\maketitle
	
	\section{Introduction}
	Nonreciprocal responses become hot topics in condensed matter physics now ~\cite{TokuraNagaosa}. 
	It is generally expected that the response 
	to the external field is different from that of the field in the opposite direction in the presence of  broken inversion symmetry $\cal{P}$.  
	When the flow of electrons, i.e., current, is concerned, the reversal of the arrow of time, i.e., the time-reversal symmetry $\cal{T}$ is also relevant, and it often happens that the nonreciprocal transport occurs when both $\cal{P}$ and 
	$\cal{T}$ are broken simultaneously although only $\cal{P}$ breaking is enough in some cases. In the normal state of the conductor, 
	the typical energy scale is the Fermi energy of the order of $eV$, which is large compared with the spin-orbit interaction and Zeeman energy due to the external magnetic field, both of which are 
	needed to introduce the asymmetry of the energy band dispersion 
	$\varepsilon_n(k)$ between $k$ and $-k$. Therefore, the value of $\gamma$, which characterizes the strength 
	of the nonreciprocal resistivity in the empirical expression 
	\begin{equation}
	\rho(I) = \rho_0 ( 1 + \gamma I B),
	\label{eq:MCA}
	\end{equation}
	is usually very small typically of the order of 
	$\sim 10^{-3}-10^{-1}$A$^{-1}$ T$^{-1}$  \cite{Rikken1,Rikken2,Rikken3,Rikken4}. Here $\rho_0$ 
	is the linear resistivity without a magnetic field, $I$ is current, and $B$ is the magnetic field. This phenomenon is called magneto-chiral anisotropy (MCA). 
	It has been reported that $\gamma$ reaches the order of 1A$^{-1}$ T$^{-1}$   in BiTeBr, which shows a gigantic bulk Rashba splitting \cite{IwasaNatPhys}.
	MCA can occur also in superconductors, where the resistivity is finite 
	above  the transition temperature or due to the vortices \cite{Hoshino}. 
	Especially, the noncentrosymmetric two-dimensional 
	superconductors have been studied from this viewpoint, and the very large 
	$\gamma$-values $\sim 10^3-10^4$A$^{-1}$ T$^{-1}$ compared with the normal 
	state are realized there \cite{WakatsukiSciAdv}. 
	It is interpreted as the replacement of the 
	energy denominator from the Fermi energy to the superconducting gap energy, 
	corresponding to the difference between the fermionic and bosonic transport. 
	Some other superconductors are reported to show MCA \cite{IwasaNatC,Itahashi}. 
	
	The nonreciprocal response can be also defined without the resistivity expressed in eq.(\ref{eq:MCA}). Instead, the critical current $I_c$ can depend on the direction of the current. In ref.\cite{OnoNat}, this nonreciprocal $I_c$ was observed in an artificial superlattice [Nb/V/Ta]$_n$ 
	under an external magnetic field. The difference 
	between the magnitudes of the critical currents in the opposite directions 
	$\Delta I_c = I_c^+ - \mid I_c^{-}\mid$ is typically 0.2mA while $I_c^{+}(\mid I_{c}^{-} \mid) \cong 6$mA, which indicates that the magnitude of the nonreciprocity is of the order of a few \%. 
		Later, there are several experiments which report the larger magnitude of the 
		nonreciprocity \cite{Pal2022,Narita2022,Jeon2022,Bauriedl2022}. 
		On the other hand, theories of nonreciprocal critical current, i.e., $\Delta I_c$, have been developed recently \cite{James2022,Daido,Fu,Bergeret,Bovkova,Schrade,Chen2022,Daido2022,Kokkeler2022}. 
	Compared with bulk transport in superconductors, 
	the Josephson junction might show the much larger diode effect, 
	because the kinetic energy at the junction is suppressed and the interaction effect can be relatively enhanced. 
	In \cite{Misaki}, the asymmetric charging energy, which acts as the 
	''kinetic energy'' of the Josephson phase $\varphi$, leads to the diode effect through the nonreciprocal dynamics of $\varphi$. In this scenario, no time-reversal symmetry breaking is needed. 
On the other hand, with $\cal{T}$ breaking, the nonreciprocal current-phase relation can lead to the diode effect even without the charging energy. 
Our target system is the superconductor (S) / Ferromagnetic insulator (FI) 
	/S junction on a three-dimensional topological insulator (TI) 
	where pairing symmetry of superconductor is $d$-wave. 
One of the merit to use $d$-wave superconductor is its high 
		transition temperatures realized in high 
		$T_{C}$ cuprate. 
		The transition temperature of high $T_{C}$ 
		cuprate is ten times larger than that of 
		conventional $s$-wave superconductor used in many junctions now.  We can expect the large magnitude of Josephson current as compared to the conventional one. 
		Also, by considering the $d$-wave/FI/$d$-wave junction, we can expect large magnitude of non-reciprocity owing to the huge spin-orbit coupling on the surface of TI. 
\par
	It is known that the standard current-phase relation (CPR) 
	of Josephson current $I(\varphi)$ between two superconductors is
	$I(\varphi) \sim \sin \varphi$, where the  $\varphi$ 
	is the macroscopic phase difference between two superconductors. 
	However, if we consider unconventional superconductors 
	like $d$-wave one, a wide variety of current phase relations appears. 
	For $d$-wave superconductor junctions, 
	when the lobe direction of $d$-wave pair potential and 
	the normal to the interface is not parallel, 
	so called zero energy 
	Andreev bound state (ZEABS) is generated at the interface 
	due to the sign change of the $d$-wave pair potential on the Fermi surface \cite{Hu94,TK95,kashiwaya00,ABSR2}. 
	The presence of ZEABS enhances the $\sin 2\varphi$ component of $I(\varphi)$ 
	and the resulting free energy minimum of the junction can locate
	neither at $\varphi=0$ nor $\pm \pi$ \cite{Yip1993,TKJosephson}. 
	Also, the non-monotonic temperature dependence of Josephson current 
	is generated by ABS depending on the direction of the 
	crystal axis of $d$-wave pair potential  
	\cite{TKJosephson,TKJosephson2,YBJosephson,Testa} . \par
If we put  S/FI/S junction with $d$-wave 
superconductors on the surface of the 
TI, it is possible to  generate a $\cos\varphi$ term in Josephson current since this system can break both $\cal{P}$ and $\cal{T}$ symmetry 
due to the strong spin-orbit coupling of TI \cite{TYN09,Linder10a}, allowing for a $\cos \varphi$ harmonic (see Appendix 3). 
		Then, we can expect exotic 
		current-phase relation with 
		$I(\varphi) \neq -I(-\varphi)$ 
		\cite{LuBo2015}. 
		One of the merit to use the S/FI/S junction on TI is that the 
		$\cos\varphi$ term is easily induced even for the narrow width of FI region without suppressing the $\sin 2\varphi$ term  \cite{LuBo2015}. 
		Then, we can realize the 
		simultaneous existence of $\sin\varphi$, $\cos\varphi$ and $\sin 2\varphi$
		terms with almost the same order. 
		This condition is essential to get 
		larger values of $Q$ factor of
		the diode effect. 

		Although the previous article has not reported the 
		nonreciprocity of the Josephson current \cite{LinderPRB2010,LuBo2015}, 
		we anticipate that the positive maximum magnitude of $I(\varphi)$, $i.e.$, $I_{c}^{+}={\rm max}(I(\varphi))$ and the negative one $I_{c}^{-}$ can take 
		the different value each other by searching various configurations of the 
		junctions with breaking mirror inversion symmetry along the interface.\par
	In this paper, we calculate Josephson current in a $d$-wave superconductor ($x<0$)/ ferromagnetic insulator ($0<x<d$)/ 
	$d$-wave superconductor ($x>d$)(S/FI/S) junctions on a 3D topological 
	insulator (TI) surface.
 It is known that the ABS generated between S/FI (FI/S) interface becomes  
Majorana bound states (MBS) \cite{FK08,Linder10a}
due to the spin-momentum locking. 
We show anomalous current phase relation and the energy dispersion of  MBS. A giant diode effect with a huge quality factor $Q$ given by 
$Q=(I_{c}^{+} - \mid I_{c}^{-} \mid)/ (I_{c}^{+} + \mid I_{c}^{-} \mid)$ 
is obtained by tuning the crystal axis of $d$-wave superconductor.  
We also clarify the strong temperature dependence of 
$Q$ due to the presence of assymetric $\varphi$ dependence of MBS. 
It is revealed how the sign of $Q$ is 
controlled by the direction of the  magnetization. \par
	The organization of this paper is as follows. We explain the model 
	and formulation in section II. The detailed expressions of the 
	Andreev reflection coefficients are shown since 
	these quantities are essential to understand the current-phase relation  
	$I(\varphi)$ for various parameters. 
	Section III shows numerically obtained 
	results about $I(\varphi)$, $Q$ and dispersion of MBS. 
	In section IV, we conclude our results. 
	
	\section{Model and Formulation} 
First, we explain the outline of the 
		way to calculate 
		Furusaki-Tsukada's formalism \cite{Furusaki91}. 
		It is known that to calculate Josephson current,
		Matsubara Green's function is needed.
		However, in non-uniform superconducting systems like junctions,
		it is difficult to obtain Matsubara Green's  function directly.
		On the other hand, it is possible to calculate the retarded Green's function by using the scattering state of the wave function. 
		This method has been used to obtain Green's function in Josephson current in unconventional 
		superconductor and junctions on the surface of 
		topological insulator  \cite{TKJosephson,kashiwaya00,LuBo2018}. 
		After we obtain the analytical formula of the retarded Green's function,
		we have obtained the Matsubara Green's function by analytical continuation
		from real energy to Matsubara frequency.
		Using the resulting Matsubara Green's function analysis, we obtain the compact relation of Josephson current given by Andreev reflection coefficient which
		is analytically continued from real energy obtained in scattering state to Matsubara frequency \cite{Furusaki91}. 
\subsection{Model}
\begin{figure}[t]
		\begin{center}
			\includegraphics[width=12cm]{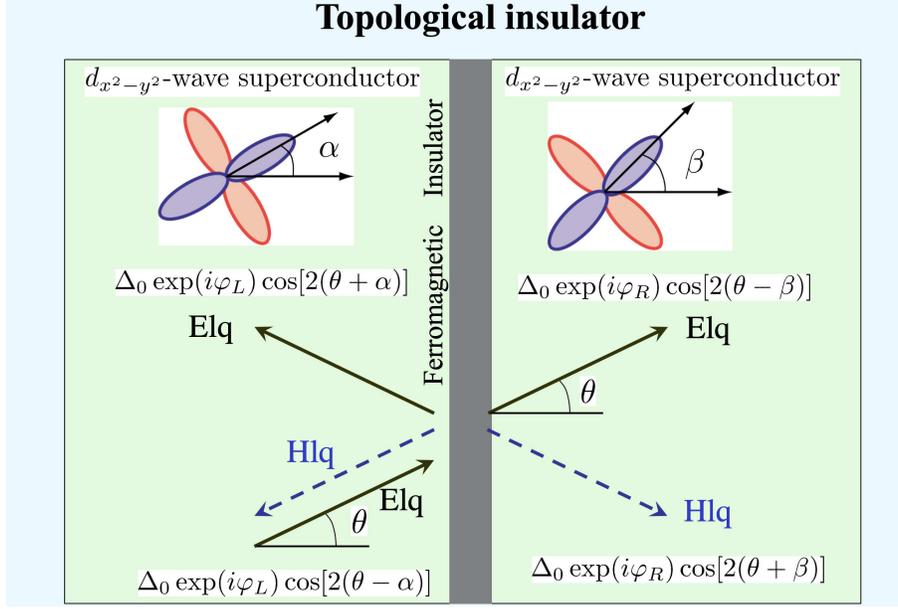}
		\end{center}
		\caption{Schematic illustration of the $d$-wave superconductor 
			junctions on the surface of a 3D topological insulator (TI). An electron-like quasiparticle (Elq) is injected and 
			it is reflected or transmitted as Elq 
			and hole-like quasiparticle (Hlq). 
				$\Delta_{L+}(\theta)=\Delta_{0}\cos\left[ 2 \left( \theta - \alpha \right) \right]\exp(i\varphi)$ and 
				$\Delta_{R+}(\theta)=\Delta_{0}\cos\left[ 2 \left( \theta - \beta \right) \right]$ are pair potentials felt by quasiparticle with the direction $\theta$, where the angle $\theta$ is measured from the normal to the interface. 
				$\Delta_{L-}(\theta)=\Delta_{0}\cos\left[ 2 \left( \theta + \alpha \right) \right]\exp(i\varphi)$ and 
				$\Delta_{R-}(\theta)=\Delta_{0}\cos\left[ 2 \left( \theta + \beta \right) \right]$ are pair potentials felt by quasiparticle with the direction 
				$\pi-\theta$, where the angle $\theta$ is 
measured from the normal to the interface.
		}
		\label{Fig1}
\end{figure}

	We consider a $d$-wave superconductor ($x<0$)/ ferromagnetic insulator ($0<x<d$)/ $d$-wave 
	superconductor ($x>d$)(S/FI/S) junction on a 3D topological insulator (TI) surface as depicted in Fig.\ref{Fig1}. 
	The corresponding Bogoliubov-de Gennes (BdG) Hamiltonian is given by 
	\cite{LuBo2015}
	\begin{equation}
	\mathcal{H}=\left[
	\begin{array}{cc}
	\hat{h}\left(k_{x},k_{y} \right)+\hat{M} & i\hat{\sigma}_{y}\Delta \left( \theta,x \right) \\
	-i\hat{\sigma}_{y}\Delta ^{\ast }\left( \theta, x \right) & -\hat{h}^{\ast
	}\left(-k_{x},-k_{y} \right)-\hat{M}^{\ast }%
	\end{array}%
	\right] ,  \label{Hamiltonian}
	\end{equation}
	with
	\[\hat{h}\left(k_{x},k_{y} \right)= 
	v\left(k_{x}\hat{\sigma}_{x} + 
	k_{y}\hat{\sigma}_{y} \right)
	-\mu \left[\Theta \left( -x\right) +\Theta \left( x-d\right) \right], 
	k_{x}=\frac{\partial}{i \partial x}, 
	\ \ k_{y}=\frac{\partial}{i \partial y}\]
	where $\hat{\sigma}_{x,y,z}$ is the Pauli matrix  in the spin space with $\hbar=1$ unit. $\mu $ 
	is the chemical potential in the superconducting  region with $\mu=vk_{F}$ and $(x,y)$ 
	component of the Fermi momentum $k_{F}$ is given by  
	$(k_{Fx},k_{Fy})=k_{F}(\cos\theta, \sin\theta)$ with an injection angle $\theta$. A chemical potential in the FI is set to be zero and  
	an exchange field in the FI region is given by 
	\cite{TYN09}
	\[
	\hat{M}=m_{z}\hat{\sigma}_{z}
	\Theta \left( x\right) \Theta \left( d-x\right) 
	\]
	and a pair potential of $d$-wave superconductor 
	is expressed by \cite{TKJosephson} 
	\begin{equation}
	\label{2.3}
	\Delta(\theta, x)
	=
	\left\{\begin{array} {ll}
	\Delta_{L\pm}\left( \theta \right) = 
	\Delta_{0}\cos\left[ 2 \left( \theta \mp \alpha \right) \right]\exp(i\varphi), & x<0 \\
	\Delta_{R\pm}\left( \theta \right)
	=\Delta_{0}\cos\left[ 2 \left( \theta \mp \beta \right) \right], & x>d. 	\end{array}
	\right.
	\end{equation}
	
	Here, $\Delta_{0}$ is a real number and 
	its temperature dependence is determined by mean field approximation 
	\cite{TKJosephson, TKJosephson2}. 
	$\alpha$ and $\beta$ denote angles 
	between the $x$-axis and the lobe direction of the pair potential of 
	the $d$-wave superconductor as shown in Fig. \ref{Fig1}. 
The index $+$ ($-$) 
in $\Delta_{L\pm}(\theta)$ and $\Delta_{R\pm}(\theta)$  
denotes the direction of the quasiparticle with the angle $\theta$ 
($\pi - \theta$) measured from the normal to the interface. 
\subsection{Wave functions of BdG equation}
	A BdG wave function of the above Hamiltonian is given by 
	\[
	\Psi\left(\bm{x}\right)=
	\exp\left(ik_{Fy}y\right)\left[\Psi_{SL}\left(x\right)\Theta\left(-x\right) + 
	\Psi_{FI}\left(x\right) \Theta\left(x\right)\Theta\left(d-x\right) + \Psi_{SR}\left(x\right) \Theta\left(x-d\right)\right]
	\] 
	with the momentum parallel to the interface $k_{Fy}$. 
	We denote the quasiparticle energy measured from the 
	Fermi surface as $E$ and 
	assume the conditions 
	where
	$\mid E \mid \ll \mu$, 
	$\Delta_{0} \ll \mu$, 
	$\mid E \mid \ll \mid m_{z} \mid$, and
	$\Delta_{0}  \ll \mid m_{z} \mid$ 
	are satisfied. 
	If we consider an electron-like quasiparticle injection from the 
	left superconductor, 
	$\Psi_{SL}(x)$, $\Psi_{FI}(x)$, and $\Psi_{SR}(x)$ 
	are given by 
	\begin{eqnarray}
	\Psi_{SL}\left(x\right) &=& 
	\left(\Psi^{e}_{in} \!+\! a_{e} \Psi_{hr}\right)\exp\left(ik_{Fx}x\right) 
	+ b_{e} \Psi_{er} \exp\left(-ik_{Fx}x\right),
	\label{wave1}
	\\
	\Psi_{FI}\left(x\right) &=& 
	f_{1e} \Psi_{e1}\exp\left(-\kappa_{ex} x\right) 
	+ f_{2e} \Psi_{e2}\exp\left(\kappa_{ex} x\right) 
	\nonumber
	\\
	&&+ f_{3e} \Psi_{h1}\exp\left(\kappa_{hx}x\right) 
	+ f_{4e} \Psi_{h2}\exp\left(-\kappa_{hx}x\right),
	\label{wave2}
	\\
	\Psi_{SR}\left(x\right) &=& 
	c_{e} \Psi_{et} \exp\left(ik_{Fx}x\right) 
	+ d_{e} \Psi_{ht} \exp\left(-ik_{Fx}x\right),
	\label{wave3}
	\end{eqnarray}
	\[
	k_{Fx}=\sqrt{\left(\mu/v\right)^{2}-k_{Fy}^{2}}, \ \ 
	\kappa_{ex}=\kappa_{hx}=\sqrt{m_{z}^{2} + v^{2} k_{y}^{2}}/v.
	\]
	$\Psi^{e}_{in}$, $\Psi_{hr}$, $\Psi_{er}$ 
	defined in the left superconductor 
	are given by
	\begin{equation}
	\Psi^{e}_{in} =
	\begin{pmatrix}
	1 \\ 
	\exp\left(i\theta\right) \\
	-\Gamma_{L+}
	\exp\left[i \left(\theta - \varphi \right)\right] \\
	\Gamma_{L+} \exp\left(-i\varphi \right)
	\\
	\end{pmatrix}, 
	\ \ 
	\Psi_{hr} =
	\begin{pmatrix}
	\Gamma_{L+} \\ 
	\Gamma_{L+}\exp\left(i\theta \right) \\ 
	-\exp\left[i \left(\theta -\varphi \right)\right] \\
	\exp\left(-i\varphi \right)
	\end{pmatrix}, 
	\ \ 
	\Psi_{er} =
	\begin{pmatrix}
	1 \\
	-\exp\left(-i\theta\right) \\
	\Gamma_{L-} \exp\left[-i \left(\theta + \varphi \right)\right] 
	\\
	\Gamma_{L-} \exp\left(-i\varphi \right)
	\end{pmatrix} 
	\end{equation}
	with $\exp(i\theta)=(k_{Fx}+ ik_{Fy})/k_{F}$. 
	%
	%
	$\Psi_{e1}$, 
	$\Psi_{e2}$, 
	$\Psi_{h1}$, and 
	$\Psi_{h2}$ in FI are 
	\begin{equation}
	\Psi_{e1} =
	\begin{pmatrix}
	i\gamma\\ 
	1 \\
	0 \\
	0 
	\end{pmatrix}, \ \ 
	\Psi_{e2} =
	\begin{pmatrix}
	-i\gamma^{-1} \\
	1 \\
	0 \\
	0 
	\end{pmatrix}, 
	\Psi_{h1} =
	\begin{pmatrix}
	0 \\
	0 \\
	i\gamma \\
	1
	\end{pmatrix}, 
	\Psi_{h2} =
	\begin{pmatrix}
	0 \\
	0 \\
	-i\gamma^{-1} \\ 
	1 
	\end{pmatrix}
	\end{equation}
	$\gamma=-v (\kappa_{ex} - k_{Fy})/m_{z}$.  
	$\Psi_{et}$, $\Psi_{ht}$ in the right superconductor are given by 
	\begin{equation}
	\Psi_{et} =
	\begin{pmatrix}
	1 \\
	\exp\left(i\theta\right) \\
	-\Gamma_{R+} \exp\left(i\theta\right) \\
	\Gamma_{R+} 
	\end{pmatrix}, \ \ 
	\Psi_{ht} =
	\begin{pmatrix}
	\Gamma_{R-} \\
	-\Gamma_{R-} \exp\left(-i \theta\right) \\
	\exp\left(-i\theta \right) \\
	1 
	\end{pmatrix}
	\end{equation}
	with
	\[
	\Gamma_{L\pm}
	=\frac{\Delta_{L\pm}\left(\theta\right)}{E + 
		\sqrt{E^{2} -\Delta^{2}_{L\pm}\left(\theta\right)}}, \ 
	\Gamma_{R\pm}
	=\frac{\Delta_{R\pm}\left(\theta\right)}{E + 
		\sqrt{E^{2} -\Delta^{2}_{R\pm}\left(\theta\right)}}. 
	\]
	We can also calculate the 
	wave function corresponding to eqs. (\ref{wave1}), (\ref{wave2}), and 
	(\ref{wave3}) with hole-like quasiparticle injection as follows.
	\begin{eqnarray}
	\Psi_{SL}\left(x\right) &=& 
	\left(\Psi^{h}_{in} \!+\! a_{h} \Psi_{er}\right)\exp\left(-ik_{Fx}x\right) 
	+ b_{h} \Psi_{hr} \exp\left(ik_{Fx}x\right), 
	\label{waveh1}
	\\
	\Psi_{FI}\left(x\right) &=& 
	f_{1h} \Psi_{e1}\exp\left(-\kappa_{ex}x\right) 
	+ f_{2h} \Psi_{e2}\exp\left(\kappa_{ex}x\right)  
	\nonumber
	\\
	&&+ f_{3h} \Psi_{h1}\exp\left(\kappa_{hx}x\right) 
	+ f_{4h} \Psi_{h2}\exp\left(-\kappa_{hx}x\right), 
	\label{waveh2}
	\\
	\Psi_{SR}\left(x\right) &=& 
	c_{h} \Psi_{et} \exp\left(ik_{Fx}x\right) 
	+ d_{h} \Psi_{ht} \exp\left(-ik_{Fx}x\right), 
	\label{waveh3}
	\end{eqnarray}
	with 
	\[
	\Psi^{h}_{in} =
	\begin{pmatrix}
	\Gamma_{L-} \\ 
	-\Gamma_{L-}\exp\left(-i\theta \right) \\
	\exp\left[-i \left(\theta + \varphi \right)\right] \\
	\exp\left(-i\varphi \right)
	\\
	\end{pmatrix}. 
	\]
	$\Psi_{SL}(x)$, $\Psi_{FI}(x)$, and $\Psi_{SR}(x)$
	satisfy the boundary conditions 
	$\Psi_{SL}(x=0)=\Psi_{FI}(x=0)$ and $\Psi_{FI}(x=d)=\Psi_{SR}(x=d)$.   
	The Andreev reflection coefficients $a_{e}$ and $a_{h}$ 
	are needed to calculate Josephson current \cite{Furusaki91,TKJosephson}. 
	They are given by 
	\begin{equation}
	a_{e}=
	-\frac{ \sigma_{N} \Lambda_{1e} + \left(1 - \sigma_{N} \right)
		\Lambda_{2e}}
	{\Lambda_{d}(E,\theta)}, \ \ 
	a_{h}=
	-\frac{ \sigma_{N} \Lambda_{1h} + \left(1 - \sigma_{N} \right)
		\Lambda_{2h}}
	{\Lambda_{d}(E,\theta)}
	\label{Andreev}
	\end{equation}
	with 
	\begin{eqnarray}
	\Lambda_{d} \left(E, \theta \right)
	&=&
	\left[1 - \sigma_{N} \right]
	\left[1 + \exp\left(-i\eta \right) \Gamma_{R+} \Gamma_{R-} \right]
	\left[1 + \exp\left(i\eta \right) \Gamma_{L+} \Gamma_{L-} \right]
	\nonumber
	\\
	&+& 
	\sigma_{N}
	\left[1 - \exp\left(-i\varphi \right) \Gamma_{L-} \Gamma_{R-} \right]
	\left[1 - \exp\left(i\varphi \right) \Gamma_{L+} \Gamma_{R+} \right]
	\label{Lambdaenergy}
	\end{eqnarray}
	\begin{eqnarray}
	\Lambda_{1e}&=&
	\left[1 - \exp(-i\varphi)\Gamma_{L-}\Gamma_{R-} \right]
	\left[\Gamma_{L+}-\Gamma_{R+}\exp\left(i\varphi\right)\right]
	\nonumber
	\\
	\Lambda_{2e}&=&
	\left[1 + \exp(-i\eta)\Gamma_{R+}\Gamma_{R-} \right]
	\left[\Gamma_{L+} + \exp\left(i\eta \right)\Gamma_{L-} \right]
	\end{eqnarray}
	
	\begin{eqnarray}
	\Lambda_{1h}&=&
	\left[1 - \exp(i \varphi)\Gamma_{L+}\Gamma_{R+} \right]
	\left[\Gamma_{L-}-\Gamma_{R-}\exp\left(-i\varphi \right) \right]
	\nonumber
	\\
	\Lambda_{2h}&=&
	\left[1 + \exp(-i\eta)\Gamma_{R+}\Gamma_{R-} \right]
	\left[\Gamma_{L-} + \exp(i\eta)\Gamma_{L+} \right]
	\end{eqnarray}
	and 
	\begin{equation}
	\cos \eta=
	\frac{m_{z}^{2}\cos^{2}\theta - \mu^{2}\sin^{2}\theta}
	{m_{z}^{2}\cos^{2}\theta + \mu^{2}\sin^{2}\theta}, \ \ 
	\sin \eta=
	\frac{-2m_{z}\mu \cos\theta \sin\theta}
	{m_{z}^{2}\cos^{2}\theta + \mu^{2}\sin^{2}\theta}.  
	\label{eta}
\end{equation}
	Here, $\sigma_{N}$ is the transparency of this junction in the normal state and it is given by 
	\begin{equation}
	\sigma_{N}
	=\frac{ \cos^{2} \theta}
	{{\rm cosh}^{2}\left(\kappa_{ex}d \right)\cos^{2} \theta
		+ {\rm sinh}^{2}\left( \kappa_{ex}d \right)
		\sin^{2} \theta \sin^{2}
		\left( \frac{\eta}{2} \right)}.
	\label{sineta}
	\end{equation}
\subsection{Josephson current formula based on Andreev reflection coefficients}
	Based on the Green's function of BdG equation, 
	it is known that Josephson current is expressed by 
	$a_{en}$ and  $a_{hn}$ which are obtained from the analytical continuation 
	from $E$ to $i\omega_{n}$ in $a_{e}$ and $a_{h}$  
	for conventional $s$-wave superconductor \cite{Furusaki91},  
	$d$-wave superconductor \cite{TKJosephson,TKJosephson2}, 
	and junctions on the TI \cite{LuBo2015,LuBo2018}, 
	where $\omega_{n}=2\pi k_{B}T(n+1/2)$ is the Matsubara frequency. 
	The resulting Josephson current $I(\varphi)$ is given by 
	\cite{TKJosephson,LuBo2015,LuBo2018}
	\begin{equation}
	R_{N}I\left( \varphi \right)
	=
	\frac{\pi \bar{R}_{N} k_{B}T}{e}
	\left\{
	\sum_{\omega_{n}}\int^{\pi/2}_{-\pi/2}
	\left[
	\frac{a_{en}\left(\theta,\varphi \right)}{\Omega_{nL+}}\Delta_{L+}\left(\theta\right)
	- 
	\frac{a_{hn}\left(\theta,\varphi \right)}{\Omega_{nL-}}\Delta_{L-}
	\left(\theta \right)
	\right]
	\cos\theta d\theta
	\right\}
	\label{JosephsonAndreev}
	\end{equation}
	with 
	\[
	\bar{R}_{N}^{-1}=\int^{\pi/2}_{-\pi/2} \sigma_{N} \cos\theta d\theta, \ 
	\Omega_{nL\pm}
	={\rm sgn}\left( \omega_{n} \right)
	\sqrt{\Delta^{2}_{L}\left(\theta_{\pm} \right) + \omega_{n}^{2}}
	\]
	and
	\begin{equation}
	a_{en}=
	i\frac{ \sigma_{N} \Lambda_{1en} + \left(1 - \sigma_{N} \right)
		\Lambda_{2en}}
	{\Lambda_{dn}(\theta,\varphi)}, \ \ 
	a_{hn}=
	i\frac{ \sigma_{N} \Lambda_{1hn} + \left(1 - \sigma_{N} \right)
		\Lambda_{2hn}}
	{\Lambda_{dn}(\theta, \varphi)}
	\end{equation}
	with 
	\begin{eqnarray}
	\Lambda_{dn} \left( \theta, \varphi \right)
	&=&
	\left[1 - \sigma_{N} \right]
	\left[1 - \exp\left(-i\eta \right) \Gamma_{nR+} \Gamma_{nR-} \right]
	\left[1 - \exp\left(i\eta \right) \Gamma_{nL+} \Gamma_{nL-} \right]
	\nonumber
	\\
	&+& 
	\sigma_{N}
	\left[1 + \exp\left(-i\varphi \right) \Gamma_{nL-} \Gamma_{nR-} \right]
	\left[1 + \exp\left(i \varphi \right) \Gamma_{nL+} \Gamma_{nR+} \right]
	\label{Lambdadn}
	\end{eqnarray}
	\begin{eqnarray}
	\Lambda_{1en}&=&
	\left[1 + \exp(-i\varphi)\Gamma_{nL-}\Gamma_{nR-} \right]
	\left[\Gamma_{nL+}-\Gamma_{nR+}\exp\left(i\varphi\right)\right]
	\nonumber
	\\
	\Lambda_{2en}&=&
	\left[1 - \exp(-i\eta)\Gamma_{nR+}\Gamma_{nR-} \right]
	\left[\Gamma_{nL+} + \exp\left(i\eta \right)\Gamma_{nL-} \right], 
	\end{eqnarray}
	
	\begin{eqnarray}
	\Lambda_{1hn}&=&
	\left[1 + \exp(i \varphi)\Gamma_{nL+}\Gamma_{nR+} \right]
	\left[\Gamma_{nL-}-\Gamma_{nR-}\exp\left(-i\varphi \right) \right]
	\nonumber
	\\
	\Lambda_{2hn}&=&
	\left[1 - \exp(-i\eta)\Gamma_{nR+}\Gamma_{nR-} \right]
	\left[\Gamma_{nL-} + \exp(i\eta)\Gamma_{nL+} \right], 
	\end{eqnarray}
	with
	\[
	\Gamma_{nL\pm}
	=\frac{\Delta_{L\pm}\left(\theta\right)}{\omega_{n} + 
		\Omega_{nL\pm}}, \ 
	\Gamma_{nR\pm}
	=\frac{\Delta_{R\pm}\left(\theta\right)}
	{\omega_{n} + \Omega_{nR\pm}}.
	\]
	By using 
	$\Gamma_{nL\pm}(\theta)=\Gamma_{nL\mp}(-\theta)$, 
	$\Gamma_{nR\pm}(\theta)=\Gamma_{nR\mp}(-\theta)$, 
	\begin{equation}
	R_{N}I(\varphi)
	=
	\frac{ \pi \bar{R}_{N}k_{B}T}{e} \sum_{n}
	\int^{\pi/2}_{-\pi/2} 
	d\theta
	\frac{4\Gamma_{nL+}\Gamma_{nR+}}
	{\mid \Lambda_{dn}
		\left( \theta, \varphi \right) \mid^{2}}
	\cos \theta \sigma_{N}
	F(\theta,i\omega_{n},\varphi)
	\label{currentphaseII}
	\end{equation}
	\begin{equation}
	F(\theta,i\omega_{n},\varphi)
	=\left( 1 - \sigma_{N} \right) \Lambda_{1n} + 
	\sigma_{N} \sin \varphi
	\mid 1 + \exp \left(i\varphi \right) \Gamma_{nL-}\Gamma_{nR-} \mid^{2}
	\label{currentphaseF}
	\end{equation}
	with 
	\begin{eqnarray}
	\Lambda_{1n}&=&
	\sin\varphi {\rm Real}
	\left[\left(1 - \exp\left(i\eta\right)\Gamma_{nL+}\Gamma_{nL-}
	\right)
	\left(1 - \exp\left(-i\eta\right)\Gamma_{nR+}\Gamma_{nR-}\right)\right]
	\nonumber
	\\
	&+&
	\cos\varphi \sin \eta
	\left(\Gamma_{nL+}\Gamma_{nL-} - \Gamma_{nR+}\Gamma_{nR-}
	\right). 
	\end{eqnarray}
	The obtained $I(\varphi)$ reproduces standard formula of $d$-wave 
	superconductor junctions without a TI 
	\cite{TKJosephson,YBJosephson,TKJosephson2,Tanaka2000} 
	by choosing $\eta=\pi$. 
	In the next section, by using eqs. (\ref{currentphaseII}) and (\ref{currentphaseF}), we calculate 
	$I(\varphi)$ and the quality factor $Q$. 
	In order to prove the $m_{z}$, $\alpha$, and $\beta$ dependence of 
	$I(\varphi)$ analytically, 
	it is convenient to transform $F(\theta,i\omega_{n},\varphi)$ 
	in eqs. (\ref{currentphaseII}) and (\ref{currentphaseF})
	as follows. 
\begin{eqnarray}
F(\theta,i\omega_{n},\varphi) 
&=&
\left( 1 - \sigma_{N} \right) 
\left( \sin\varphi\Lambda_{ne} 
	+ \cos\varphi \Lambda_{no} \right) 
\nonumber
	\\
&+&\sigma_{N} 
	\left[ \sin \varphi \left(1 + \Gamma^{2}_{nL-}\Gamma^{2}_{nR-} \right)
	+ \Gamma_{nL-}\Gamma_{nR-}\sin2\varphi \right]
	\label{formulaF}
	\end{eqnarray}
	with
	\begin{equation}
	\Lambda_{ne}=1 + \Gamma_{nL+}\Gamma_{nL-}\Gamma_{nR+}\Gamma_{nR-} 
	- \cos \eta \left(  \Gamma_{nL+}\Gamma_{nL-} + 
	\Gamma_{nR+}\Gamma_{nR-} \right), 
	\label{Lambdane}
	\end{equation}
	and 
	\begin{equation}
	\Lambda_{no}=
	\left(\Gamma_{nL+}\Gamma_{nL-} - \Gamma_{nR+}\Gamma_{nR-}
	\right)\sin \eta
	. 
	\label{Lambdano}
	\end{equation}
	Here, $\Lambda_{ne}$ and $\Lambda_{no}$ are even and odd function of 
	$\theta$, respectively. 
	
	\section{Results}
	First, let us focus on the current phase relation (CPR).
		In order to understand the obtained 
		results more intuitively, 
		we rewrite eq. (\ref{currentphaseII}) 
		as follows, 
		\begin{equation}
		R_{N}I(\varphi)
		=
		\frac{ \pi \bar{R}_{N}k_{B}T}{e} \sum_{n}
		\int^{\pi/2}_{-\pi/2} d\theta 
		\frac{2\cos\theta \sigma_{N}}
		{\mid \Lambda_{dn} 	\left( \theta,\varphi \right) \mid^{2}}
		\left[A\left(\theta\right)\sin\varphi + B\left(\theta\right)\sin2\varphi  	+ C\left(\theta \right) \cos\varphi \right]
		\label{currentphasesymmetry}
		\end{equation}
		with
		\begin{equation}
		A\left(\theta \right)
		=\left(\Gamma_{nL+}\Gamma_{nR+}+ \Gamma_{nL-}\Gamma_{nR-} \right) 
		\left[ \left(1-\sigma_{N}\right)\Lambda_{ne} + 
		\sigma_{N}\left( 1 + \Gamma_{nL+}\Gamma_{nL-}\Gamma_{nR+}\Gamma_{nR-}
		\right) \right], 
		\label{A}
		\end{equation}
		\begin{equation}
		B\left(\theta \right)
		=2\sigma_{N}\Gamma_{nL+}\Gamma_{nL-}\Gamma_{nR+}\Gamma_{nR-}, \ \ 
		C\left(\theta\right)
		=\left(1-\sigma_{N}\right)
		\left(\Gamma_{nL+}\Gamma_{nR+}-\Gamma_{nL-}\Gamma_{nR-} \right)
		\Lambda_{no}
		\label{BandC}
		\end{equation}
		using the definition of 
		$\Lambda_{dn}(\theta,\varphi)$, 
$\Lambda_{ne}$, and $\Lambda_{no}$ given in eqs. 
		(\ref{Lambdadn}), (\ref{Lambdane}) 
		and (\ref{Lambdano}). 
In general, due to the $\varphi$ dependence of $\Lambda_{dn}(\theta,\varphi)$ 
in eq. (\ref{currentphasesymmetry}), $I(\varphi)$ includes 
terms proportional to $\sin(n\varphi)$ and $\cos(n\varphi)$ with $(n \geq 1)$.
		As seen from eq.\ref{BandC}, the term which
		is proportional to $\cos \varphi$ 
		in eq. (\ref{currentphasesymmetry})
		appears when both 
		$\Lambda_{no} \neq 0$ and 
		$\Gamma_{nL+}\Gamma_{nR+} \neq  \Gamma_{nL-}\Gamma_{nR-}$ 
		are satisfied except for special $\theta$. 
		This means that 
		$\sin \eta$ in $\Lambda_{no}$ (eq. \ref{Lambdano}) 
		and $m_{z}$ in eq. (\ref{eta}) are nonzero. It is remarkable that the term proportional to $\cos \varphi$  
		in eq. (\ref{currentphasesymmetry}) 
		is induced by $m_{z}$ 
		which is in sharp contrast to the case of 
		$s$-wave superconductor Josephson junction on 
		TI where in-plane magnetic field generates 
		$\cos \varphi$ term \cite{TYN09}. 
		However, the magnitude of $m_{z}$ cannot be too large, since the coupling between two superconductors becomes weaker and
		the magnitude of $\sin(2\varphi)$ term
		is suppressed since it is basically proportional to the second order of the transparency of the junctions.
The coexistence of all three harmonics, $i.e.$, 
$\sin \varphi$, $\cos \varphi$ and $\sin 2\varphi$, 
 is essential for the Josephson diode effect. 
\par
\begin{figure}[t]
		\begin{center}
			\includegraphics[width=12cm]{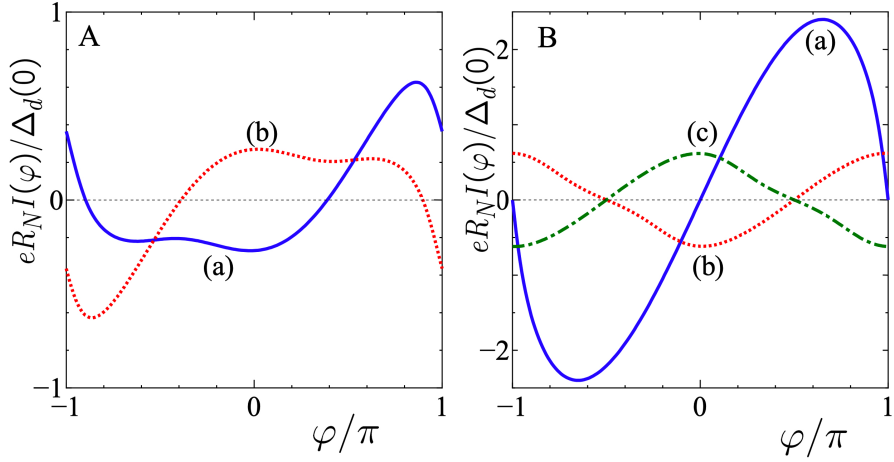}
		\end{center}
		\caption{Current phase relation $I(\varphi)$ is plotted for $T=0.05T_{d}$ and 
			$d\mid m_{z}\mid /v=1$. $R_{N}$, $\Delta_{d}(0)$ and $T_{d}$ are the resistance of the junction in the normal state, 
			the amplitude of pair potential at zero temperature, 
			and the transition temperature of $d$-wave superconductor, respectively. 
			A: $\alpha=-0.2\pi$ and $\beta=0.09\pi$. 
			(a)$m_{z}=0.5\mu$, (b)$m_{z}=-0.5\mu$. 
			B: (a)$\alpha=0$, $\beta=0$, and $m_{z}=0.5\mu$.
			(b)$\alpha=0$, $\beta=0.25\pi$ and $m_{z}=0.5\mu$. 
			(c)$\alpha=0$, $\beta=0.25\pi$ and $m_{z}=-0.5\mu$. }
		\label{Fig2}
	\end{figure}

As shown later, the quality factor $Q$ depends sensitively 
on the angles	$\alpha$ and $\beta$. 
It is noted that $\cos\varphi$ term does not appear for $C(\theta)=0$. By choosing $\eta=\pi$, we reproduce the formula of Josephson current of d-wave junctions without TI \cite{TKJosephson,TKJosephson2}. 
Here, we pick up the particular value of 
		$\alpha=-0.2\pi$ and $\beta=0.09\pi$, where $Q$ is hugely enhanced, and examine the current-phase relation. 
		In this case, all terms proportional to 
		$\sin \varphi$, $\cos \varphi$, and $\cos 2\varphi$ 
of the same order of magnitudes. 
	At this value of $\alpha$, $\beta$, we obtain quite exotic CPR shown 
	in Fig. \ref{Fig2}A. 
	
	As seen from curves (a) and (b) of Fig. \ref{Fig2}A, the magnitude of  
	$I_{c}^{+}$ and $I_{c}^{-}$ are different from each other, where 
	$I_{c}^{+}$ ($I_{c}^{-}$) is the positive (negative) maximum value of $I(\varphi)$. 
	Since the quality factor showing nonreciprocity is expressed by 
	\begin{equation}
	Q=\frac{I_{c}^{+} - \mid I_{c}^{-} \mid}{I_{c}^{+} + \mid I_{c}^{-} \mid}, 
	\end{equation}
	we can expect diode effect for nonzero $Q$. 
	On the other hand, for $\alpha=0$, $\beta=0$ 
	(curve (a) in Fig. \ref{Fig2}B)
	$I(\varphi)$ shows a standard sinusoidal behavior 
	since $\Gamma_{nL+}=\Gamma_{nL-}$, $\Gamma_{nR+}=\Gamma_{nR-}$, 
		and 
		$\Gamma_{nL+}\Gamma_{nR+}=\Gamma_{nL-}\Gamma_{nR-}$ are satisfied.  
		Then, $C(\theta)$ in  eq.(\ref{currentphasesymmetry}) becomes 
		zero and $I(\varphi=0)=I(\varphi=\pi)=0$ is consistent with curve (a) 
		in Fig \ref{Fig2}B.  
	For $\alpha=0$, $\beta=\pi/4$, although $I(\varphi)$ shows an unconventional 
	current phase relation with nonzero $I(\varphi)$ at $\varphi=0$, 
	$I_{c}^{+}=\mid I_{c}^{-} \mid$ is still satisfied 
due to the absence of 
	the term proportional to $\sin \varphi$ 
		in eq. (\ref{currentphasesymmetry}) 
		since 
		$\Gamma_{nL+}\Gamma_{nR+}+ \Gamma_{nL-}\Gamma_{nR-}=0$ is satisfied. 
		Then, $A(\theta)$ in  eq.(\ref{currentphasesymmetry})
		becomes zero and the resulting $I(\varphi=\pm \pi/2)=0$ is  
		consistent with curves (b) and (c) 
		in Fig \ref{Fig2}B. \par
By changing the sign of the magnetization from $m_{z}$ to $-m_{z}$,
	$I(\varphi)=I(\varphi,m_{z})$ satisfies
	\begin{equation}
	I(\varphi,m_{z})=-I(-\varphi,-m_{z})
	\label{currentinversion}
	\end{equation}
	as seen from curves (a) and (b) in Fig. \ref{Fig2}A
	and curves (b) and (c) in Fig. \ref{Fig2}B. 
This property can be understood from the 
time reversal operation. 
Actually, we can show this relation explicitly 
in the Appendix 1. 
	Next, we show the $\alpha$ and $\beta$ dependence of 
	$Q$ for $-\pi/4 \leq \alpha \leq \pi/4$ and 
	$-\pi/4 \leq \beta \leq \pi/4$. 
	\begin{figure}[t]
		\begin{center}
			\includegraphics[width=12cm]{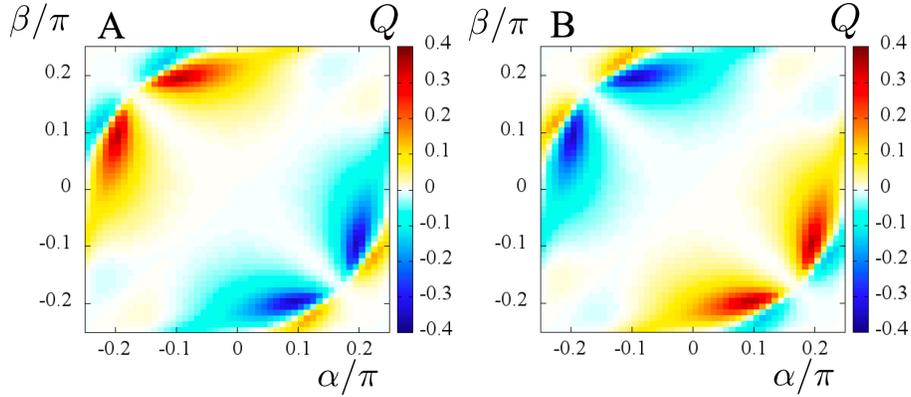}
		\end{center}
		\caption{$Q$ is plotted for various $\alpha$ and $\beta$ 
			for $T=0.05T_{d}$ and $d \mid m_{z} \mid/v=1$. 
			(a)$m_{z}=0.5\mu$, (b)$m_{z}=-0.5\mu$. 
		}
		\label{Fig3}
	\end{figure}
	It is remarkable that the maximum value of $\mid Q \mid$ becomes 
	almost 0.4 and it means the generation of the
	giant diode effect by tuning $\alpha$ and $\beta$. 
	
Here, by changing $(\alpha,\beta)$ to $(-\alpha,-\beta)$, $Q=Q(\alpha,\beta)$  
		satisfies 
		\begin{equation}
		Q\left(\alpha,\beta \right)=-Q\left(-\alpha,-\beta \right). 
		\label{Qinversion}
		\end{equation}
		We can show this relation analytically as shown in Appendix 2. 
		Also, it can be explained by more intuitive discussion. 
		If we denote the macroscopic phase by $\varphi _{L}$ and 
$\varphi_{R}$ with $\varphi =\varphi_{L}-\varphi _{R}$ 
		(we set $\varphi_{L}=\varphi$ and $\varphi_{R}=0$ in this model without loosing generality), we have
		\begin{equation}
		I\left( \varphi ,\alpha ,\beta \right) =I\left( \varphi _{1},\alpha ;\varphi
		_{2},\beta \right) ,
		\end{equation}%
		where the left superconductor has parameters $\left( \varphi _{1},\alpha
		\right) $ and the right superconductor has $\left( \varphi _{2},\beta
		\right) $. If we apply a mirror operation with respect to the $yz$ plane,
		the left superconductor has parameters $\left( \varphi _{2},\beta \right) $
		and the right superconductors has $\left( \varphi _{1},\alpha \right) $.
		Because the direction of the current reverses according to this operation,
		we have
		\begin{equation}
		I\left( \varphi _{2},\beta ;\varphi _{1},\alpha \right) =-I\left( \varphi
		_{1},\alpha ;\varphi _{2},\beta \right) .
		\end{equation}%
		Therefore, we have%
		\begin{equation}
		I\left(\varphi, \alpha,\beta \right)=-I\left( -\varphi,\beta,\alpha
		\right).
		\end{equation}
		This relation leads to eq. (\ref{Qinversion}). \par
\begin{figure}[t]
		\begin{center}
			\includegraphics[width=12cm]{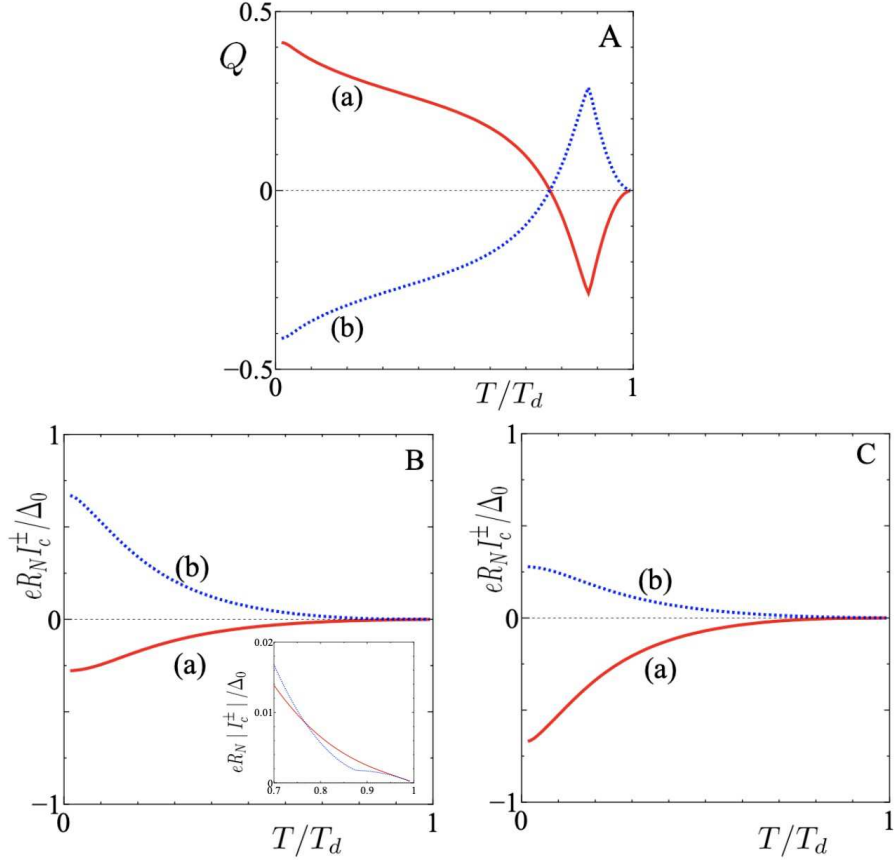}
		\end{center}
		\caption{Temperature dependences of $Q$, 
			$I_{c}^{+}$ and $I_{c}^{-}$ are plotted 
			for  $\alpha=-0.2\pi$, $\beta=0.09\pi$ and 
			$d \mid m_{z} \mid/v=1$. 
			A: $Q$ for (a)$m_{z}=0.5\mu$, (b)$m_{z}=-0.5\mu$. 
			B: (a)$I_{c}^{-}$ and (b)$I_{c}^{+}$ for $m_{z}=0.5\mu$. In the inset, $\mid I_{c}^{\pm} \mid$ is plotted for $0.7T_{d} < T < T_{d}$. C: (a)$I_{c}^{-}$ and (b)$I_{c}^{+}$ for $m_{z}=-0.5\mu$.	}
		\label{Fig4}
	\end{figure}

It is interesting to clarify how nonreciprocal effect depends on the temperature. 
	As shown in Fig. \ref{Fig4},  $Q$ is enhanced at low temperatures and 
	has a sign change at $T=T_{p}$ with $T_{p} \sim 0.78T_{d}$. 
	Also, there is a sharp peak structure of $Q$ at $T = 0.85T_d$. 
		This peak structure comes from the 
		intrinsic nature of temperature dependence of $d$-wave superconductor junctions. In $d$-wave superconductor junctions, if we consider injection angle resolved 
		Josephson current, we can decompose into $0$-junction and $\pi$-junction 
		domains. The temperature dependence of Josephson current from
		$0$-junction domain and that of $\pi$-junction domain
		can be qualitatively very different shown in previous papers
		\cite{TKJosephson2,kashiwaya00}. 
		Then, the macroscopic phase difference
		$\varphi=\varphi_{m}$. which gives a maximum Josephson current has a
		jump at some temperature.
		The resulting maximum Josephson current has a kink like
		structure as shown in Figs. 36 and 37 in Ref. \cite{kashiwaya00}. 
		This is the reason why $Q$ has a sharp peak
		at $T \simeq 0.85T_d$. \par
	As shown in curves (a) and (b) in Fig.\ref{Fig4}A, 
	the overall sign of $Q$ is reversed with the sign change of $m_{z}$. 
	The corresponding $I_{c}^{-}$ and $I_{c}^{+}$ are plotted 
	as curves (a) and (b) for $m_{z}=0.5\mu$ in Fig.\ref{Fig4}B and 
	those for $m_{z}=-0.5\mu$ in Fig.\ref{Fig4}C. 
	If we denote $m_{z}$ dependence of $I_{c}^{\pm}$ explicitly, 
	$I_{c}^{\mp}(m_{z}=0.5\mu)=-I_{c}^{\pm}(m_{z}=-0.5\mu)$ 
	to be consistent with eq. (\ref{currentinversion}). 
	In the inset of Fig.\ref{Fig4}B, 
	$\mid I_{c}^{\pm}\mid$ is plotted in the enlarged scale from 
	$0.7T_{d} < T < T_{d}$. 
	$I_{c}^{+}=\mid I_{c}^{-} \mid$ is satisfied for $T=T_{p}$ 
	when $Q$ becomes zero as shown in Fig. \ref{Fig4}B.  \par
	To elucidate the exotic CPR specific to nonreciprocal nature of Josephson current, we focus on its Fourier components. 
	In general, Josephson current is decomposed into 
	\begin{equation}
	I(\varphi) = \sum_{n} \left[ I_{n} \sin n\varphi
	+ J_{n} \cos n\varphi \right]. 
	\label{currentphase}
	\end{equation}
	\begin{figure}[t]
		\begin{center}
			\includegraphics[width=12cm]{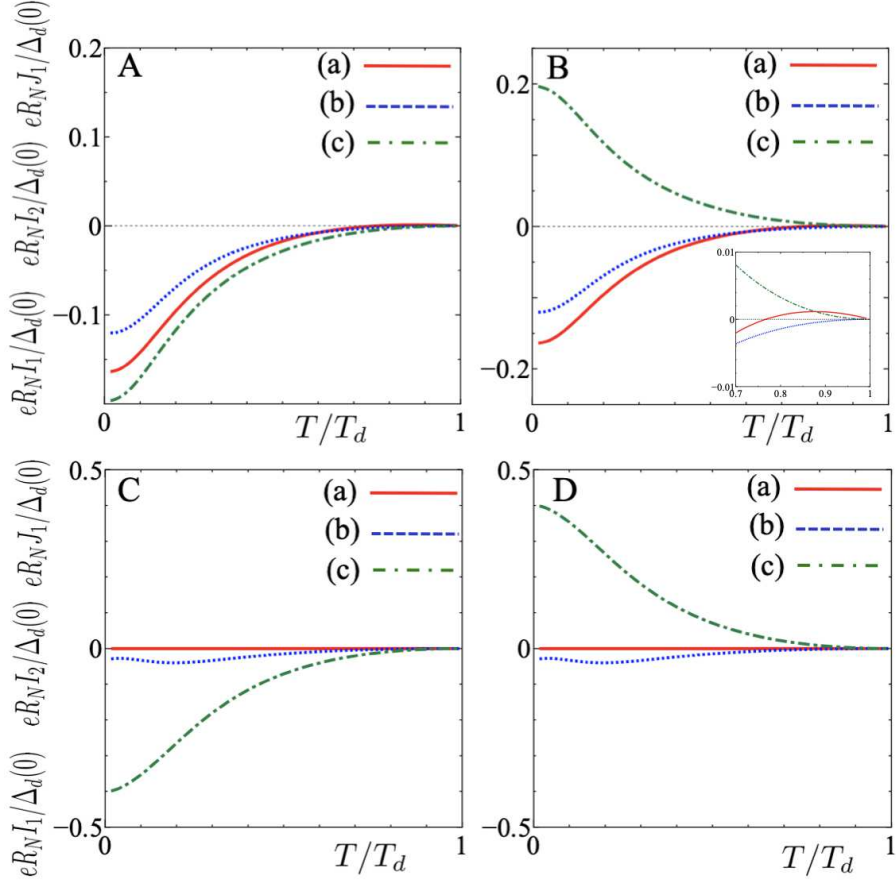}
		\end{center}
		\caption{
			Temperature dependences of (a)$I_{1}$, (b)$I_{2}$ and 
			(c)$J_{1}$ are plotted for $d \mid m_{z} \mid/v=1$. 
			A: $(\alpha,\beta)=(-0.2\pi, 0.09\pi)$ with  $m_{z}=0.5\mu$, 
			B: $(\alpha,\beta)=(-0.2\pi, 0.09\pi)$ with $m_{z}=-0.5\mu$,
			C: $(\alpha,\beta)=(0, 0.25\pi)$ with $m_{z}=0.5\mu$, 
			and D: $(\alpha,\beta)=(0, 0.25\pi)$ with $m_{z}=-0.5\mu$. 
			The enlarged plot for $0.7T_{d} < T <T_{d}$ is shown in B as  the inset. }
		\label{Fig5}
	\end{figure}
	For $\alpha=-0.2\pi$ and $\beta=0.09\pi$, 
	$I_{1}$, $I_{2}$, and $J_{1}$ become nonzero values 
	(Figs. \ref{Fig5}A and B). By changing $m_{z}$ to $-m_{z}$, 
	$I_{1}$ and $I_{2}$ are invariant and $J_{1}$ has the sign change as shown in 
	Figs. \ref{Fig5}A and B. 
	As shown in the inset of Fig. \ref{Fig5}B,  $I_{1}$ has the sign change at 
	$T=T_{p}$. At this temperature, as shown in Fig. \ref{Fig4}A, $Q$ becomes zero. 
	We also show $I_{1}$, $I_{2}$, and $J_{1}$ 
	for $\alpha=0$ and $\beta=0.25\pi$ in Figs. \ref{Fig5}C and D. 
	In this case, the resulting $Q$ is zero since 
	the term proportional to $A=A(\theta)$ in 
	eq. (\ref{currentphasesymmetry}) becomes zero, and 
	the resulting $I_{1}$ becomes zero independent of the sign of $m_{z}$. 
	
	Similar to the case for Figs. \ref{Fig5}A and B, 
	$I_{2}$ is invariant and $J_{1}$ has a sign change 
	by changing $m_{z}$ to $-m_{z}$. 
	To summarize, the simultaneous existence of 
	$I_{1}$, $I_{2}$ and $J_{1}$ does lead  to  nonzero $Q$.  
	%

	
	
	Next, we discuss the energy spectrum of the ABSs since 
	it plays a crucial role to determine $I(\varphi)$ 
	\cite{FT1991b,Beenakker91,kashiwaya00,ABSR2,Yakovenko,TK96}. 
It is known that the magnitude of $I(\varphi)$ is enhanced at low temperatures
		due to the presence of low energy ABS.
	In addition, by  the strong spin-momentum locking of the surface states of topological insulator (TI), the ABSs
	in the present S/FI/S junction become MBSs 
	\cite{FK08,ANB09,LLN09,TYN09,Linder10a}. 
The non-reciprocity, which is responsible for the diode effect, is also 
apparent in the spectrum of the ABS in the junction. 
\par
	The energy eigenvalues of ABS(MBS) $E_{b}$ are 
	found by the zero of $\Lambda_{d}(E,\theta)$ defined in 
	eq. (\ref{Lambdaenergy}) for 
	\begin{equation}
	\mid E_{b} \mid < {\rm min}\left( \mid \Delta_{L+} \mid, 
	\mid \Delta_{L-} \mid, \mid \Delta_{R+} \mid, 
	\mid \Delta_{R-} \mid \right). 
	\label{boundstatecondition}
	\end{equation}
	Only for limited cases,  we can obtain the energy level of $E_{b}$ 
	analytically. 
	For $\alpha=\beta=0$, the energy level of the ABS is expressed by 
	\begin{equation}
	E_{b}=\pm \sqrt{ \sigma_{N} \cos^{2} \frac{\varphi}{2} + 
		\left(1 - \sigma_{N} \right) \sin^{2} \frac{\eta}{2} }
	\mid \cos 2\theta \mid \Delta_{0}
	\label{Majoranaboundstate1}
	\end{equation}
	with 
	\[
	\cos^{2}\frac{\eta}{2}=\frac{m_{z}^{2}\cos^{2}\theta}
	{m_{z}^{2}\cos^{2}\theta + \mu^{2} \sin^{2}\theta}, \ \ \
	\sin^{2}\frac{\eta}{2}=\frac{\mu^{2}\sin^{2}\theta}
	{m_{z}^{2}\cos^{2}\theta + \mu^{2} \sin^{2}\theta}
	\]
	to be consistent with the result of an 
	$s$-wave superconductor junction \cite{TYN09}. 
	$E_{b}$ becomes zero for $\varphi=\pm \pi$ and $\theta=0$. 
\par
For $\alpha=\beta=\pi/4$, 
	$E_{b}$ becomes 
	\begin{equation}
	E_{b}=\pm \sqrt{ \sigma_{N} \cos^{2} \frac{\varphi}{2} + 
		\left(1 - \sigma_{N} \right) \cos^{2} \frac{\eta}{2} }
	\mid \sin 2\theta \mid \Delta_{0}. 
	\label{Majoranaboundstate2}
	\end{equation}
	$E_{b}$ is zero for $\varphi=\pm \pi$ and $\theta=\pm \pi/2$ 
	or $\varphi=\pm \pi$ and $\theta=0$. 
	In this case, the pair potential also becomes zero and $E_{b}$ is 
	absorbed into the continuum level. 
 In these two cases with eqs. (\ref{Majoranaboundstate1}), and 
(\ref{Majoranaboundstate2}), 
since $E_{b}$ is a symmetric function of $\varphi$, we
can not expect diode effect and resulting $Q$ is zero. 
\par
In other cases, only for $\theta=0$ and $\varphi=\pi$, we can show $E_{b}=0$ 
	for wide variety of parameters with $-\pi/4< \alpha < \pi/4$ and 
	$-\pi/4 < \beta < \pi/4$. 
	In this case, 
	$\Gamma_{L+}=\Gamma_{L-}=\Gamma_{L}$ 
	and $\Gamma_{R+}=\Gamma_{R-}=\Gamma_{R}$ are satisfied. 
	Then, $\Lambda_{d}(E,\theta)$ becomes
	\begin{equation}
	\Lambda_{d}\left(E, \theta=0 \right)
	=\left( 1-\sigma_{N} \right)
	\left( 1 + \Gamma^{2}_{R} \right)
	\left( 1 + \Gamma^{2}_{L} \right)
	+ \sigma_{N}
	\left( 1 + \Gamma_{L} \Gamma_{R} \right)^{2}. 
	\end{equation}
	Since $\cos(2\alpha)$ and $\cos(2\beta)$ become positive numbers,  
	$\Gamma_{R}$ and $\Gamma_{L}$ become $-i$ at $E=0$  
	and $\Lambda_{d}(E,\theta)=0$ at this condition. 
	This means $E_{b}=0$ and the ubiquitous presence of the zero energy 
	ABS for various $\alpha$ and $\beta$  
	at $\varphi=\pm \pi$ and $\theta=0$. \par
\begin{figure}[t]
		\begin{center}
			\includegraphics[width=16cm]{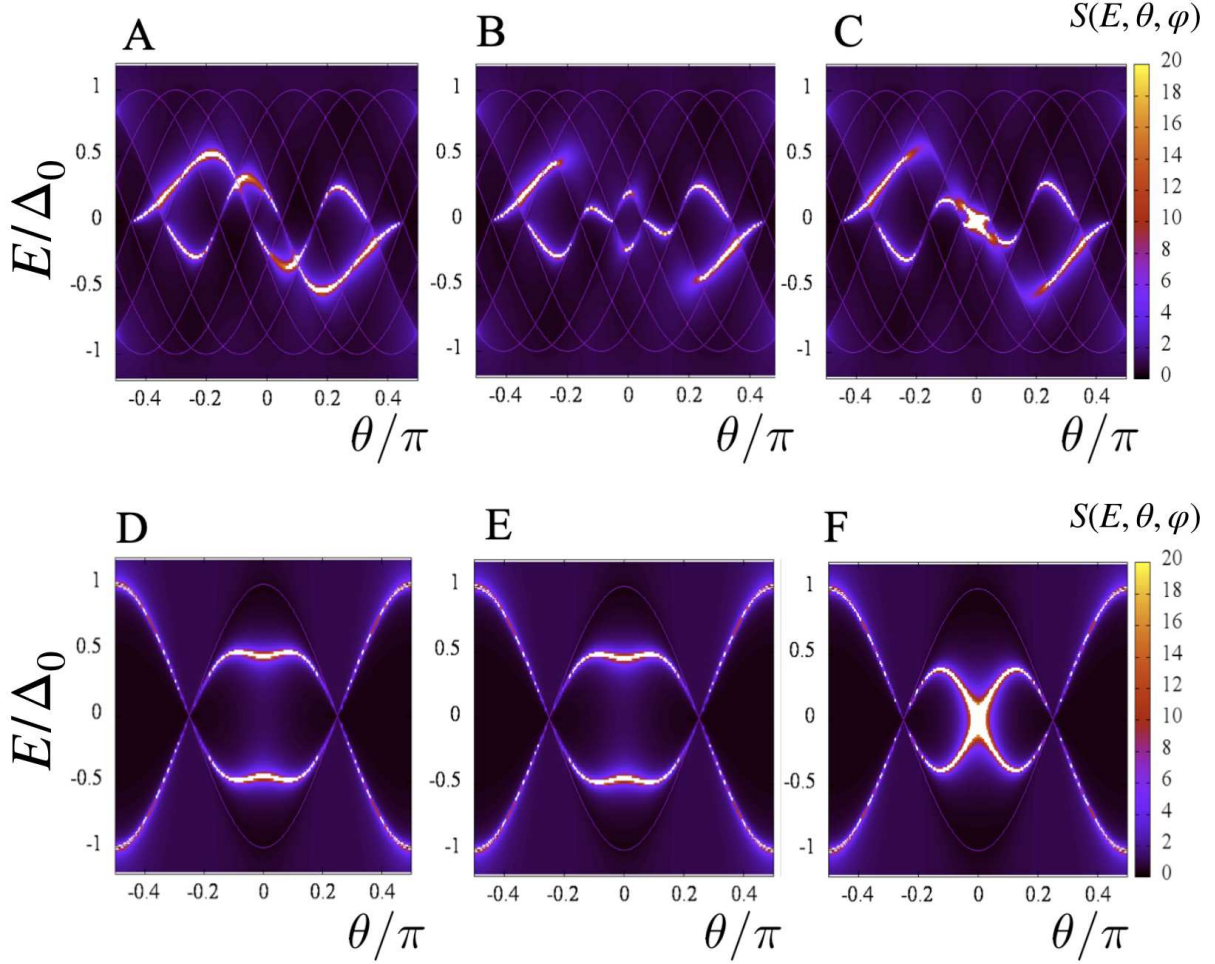}
		\end{center}
		\caption{
			The intensity plot of $S(E,\theta,\varphi)$ for 
			fixed $\varphi$ for $d \mid m_{z} \mid/v=1$ and $m_{z}=0.5\mu$. 
			$(\alpha,\beta)=(-0.2\pi, 0.09\pi)$ for A, B and C.
			$(\alpha,\beta)=(0, 0.25\pi)$ for  D, E and F. 
			A: $\varphi=0.5\pi$, B: $\varphi=-0.5\pi$, and 
			C:  $\varphi=\pi$. D: $\varphi=0.5\pi$, E: $\varphi=-0.5\pi$, and F: $\varphi=\pi$. We plot $\pm \Delta_{0}\cos[2(\theta \pm \alpha)]$ and 
			$\pm \Delta_{0}\cos[2(\theta \pm \beta)]$ as auxiliary lines. 	}
		\label{Fig6}
	\end{figure}

 
	In general, it is impossible to solve $E_{b}$ analytically, and
	we plot inverse of $\Lambda_{d}(E,\theta)= \Lambda_{d}(E,\theta, \varphi)$
	\begin{equation}
	S(E,\theta, \varphi)=\frac{1}{ \mid \Lambda_{d} 
		\left( E, \theta, \varphi \right) \mid}. 
	\end{equation}
	The intensity plot of $S(E,\theta,\varphi)$ for 
	fixed $\varphi$ is shown in Fig. \ref{Fig6}. 
	In the actual calculation we replace 
	$E$ with $E + i \delta$ 
	with a small number $\delta=0.001\Delta_{0}$ 
	to avoid the divergence, where we have used 
	the value of $\Delta_{0}$ at zero temperature.  
	
	We first show the contour plot of $S(E,\theta,\varphi)$ for fixed value of $\varphi$. 	The blight curve satisfying eq. (\ref{boundstatecondition}) 
	corresponds to the position of $E_{b}$. 
	As shown in Fig. \ref{Fig6}A, 
	$S(E,\theta,\varphi)$ shows a complicated $\theta$ dependence for 
	$\alpha=-0.2\pi$ and $\beta=0.09\pi$ 
	where nonreciprocal effect is prominent as discussed in Figs. 
	\ref{Fig2}, \ref{Fig3} and \ref{Fig4}. 
	By changing $\varphi=0.5\pi$ to $-0.5\pi$, $S(E,\theta,\varphi)$ 
	shows a dramatically different behavior as shown in Fig.\ref{Fig6}B 
	as compared to that in Fig.\ref{Fig6}A. 
From Figs. \ref{Fig6}A and \ref{Fig6}B,  we see
that the ABS energy spectrum
is different for the phase biases $\varphi$ ad $-\varphi$ 
in the regime of the Josephson diode effect.
	For $\varphi=\pi$, $S(E=0,\theta,\varphi)$ is enhanced around 
	$\theta=0$ (Fig.\ref{Fig6}C) 
	due to the existence of ABS at $E=0$. 
	For all cases (Figs.\ref{Fig6}A, \ref{Fig6}B and \ref{Fig6}C), 
	\begin{equation}
	S(E,\theta,\varphi) \neq S(E,-\theta,\varphi)
	\end{equation}
	is satisfied. 
	
	On the other hand, for $\alpha=\beta=0$, $S(E,\theta,\varphi)$ shows 
	a symmetric function with  $\theta$ (Figs. \ref{Fig6}D, E and F) 
	\begin{equation}
	S(E,\theta,\varphi)=S(E,-\theta,\varphi), \ 
	S(E,\theta,\varphi)=S(E,\theta,-\varphi)
	\end{equation}
	to be consistent with eq. (\ref{Majoranaboundstate1}). \par
	
	\begin{figure}[t]
		\begin{center}
			\includegraphics[width=14cm]{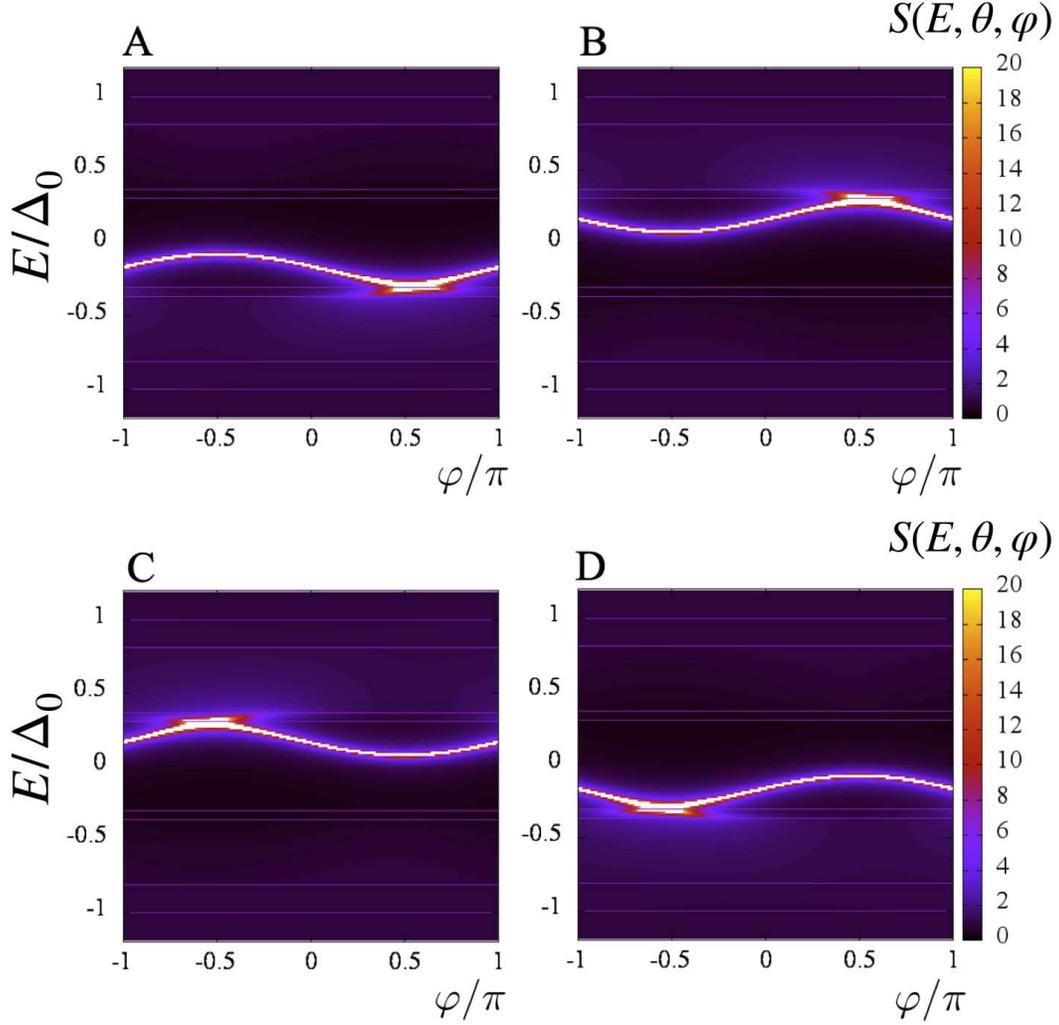}
		\end{center}
		\caption{
			The intensity plot of $S(E,\theta,\varphi)$ for 
			fixed $\theta$ for $d \mid m_{z} \mid/v=1$, $\alpha=-0.2\pi$ and $\beta=0.09\pi$. 
			A: $\theta=0.1\pi$ and $m_{z}=0.5\mu$, 
			B: $\theta=-0.1\pi$ and $m_{z}=0.5\mu$, 
			C: $\theta=0.1\pi$ and $m_{z}=-0.5\mu$, and 
			D: $\theta=-0.1\pi$ and $m_{z}=-0.5\mu$. 
			We plot $\pm \Delta_{0}\cos[2(\theta \pm \alpha)]$ and 
			$\pm \Delta_{0}\cos[2(\theta \pm \beta)]$ as auxiliary lines. 
		}
		\label{Fig7}
	\end{figure}
	In Fig. \ref{Fig7}, we focus on $\varphi$ dependence of 
	$S(E,\theta,\varphi)$ for fixed $\theta$ with  
	$\alpha=-0.2\pi$ and $\beta=0.09\pi$.  
	By changing $\theta$ to $-\theta$, $S(E,\theta,\varphi)$ has a dramatic change. 
	ABS is located for $E<0$ for $\theta=0.1\pi$ 
	while it is located for $E>0$ for $\theta=-0.1\pi$ 
	(Figs. \ref{Fig7}A and B). 
	On the other hand, if we change $m_{z}=0.5\mu$ to $m_{z}=-0.5\mu$, 
	ABS is located for $E>0$ for $\theta=0.1\pi$ 
	while it is located for $E<0$ for $\theta=-0.1\pi$ 
	(Figs. \ref{Fig7}C and D). 
 It is noted that the non-reciprocal current phase relation of $I(\varphi)$ in Fig. \ref{Fig2}A comes from the exotic $\varphi$ dependence of ABS  
as shown from  $S(E,\theta,\varphi)$ in Fig.\ref{Fig7}. 
Since $Q$ is determined by the maximum Josephson current,
its value can be enhanced by the
asymmetric energy spectrum of ABS for  $\varphi$ and $-\varphi$. \par
	Finally, we mention how the energy level of $E_{b}$ changes by 
	the transformation from $m_{z}$ to $-m_{z}$.
	By using the properties of $\Gamma_{R\pm}$, $\Gamma_{L\pm}$, and $\eta$,  
	$\Lambda(E,\theta,\varphi)=\Lambda(E,\theta,\varphi,m_{z})$ 
	and $S(E,\theta,\varphi)=S(E,\theta,\varphi,m_{z})$ satisfy
	\begin{eqnarray}
	\Lambda_{d}\left(E,-\theta,\varphi,-m_{z}\right)
	&=&
	\left[1 - \sigma_{N} \right]
	\left[1 + \exp\left(-i\eta \right) \Gamma_{R+} \Gamma_{R-} \right]
	\left[1 + \exp\left(i\eta \right) \Gamma_{L+} \Gamma_{L-} \right]
	\nonumber
	\\
	&+& 
	\sigma_{N}
	\left[1 - \exp\left(-i\varphi \right) \Gamma_{L+} \Gamma_{R+} \right]
	\left[1 - \exp\left(i\varphi \right) \Gamma_{L-} \Gamma_{R-} \right], 
	\end{eqnarray}
	
	\begin{equation}
	\Lambda_{d}\left(E,-\theta,-\varphi,-m_{z} \right)
	=
	\Lambda_{d}\left(E,\theta,\varphi,m_{z} \right)  
	\end{equation}
	and 
	\begin{equation}
	S\left(E,-\theta,-\varphi,-m_{z} \right)
	=
	S\left(E,\theta,\varphi,m_{z} \right). 
	\label{relationS}
	\end{equation}
	We can see eq.(\ref{relationS})  by comparing 
	Fig.\ref{Fig7}A (\ref{Fig7}B) and Fig. \ref{Fig7}D (\ref{Fig7}C). \par
\begin{figure}[t]
			\begin{center}
				\includegraphics[width=16cm]{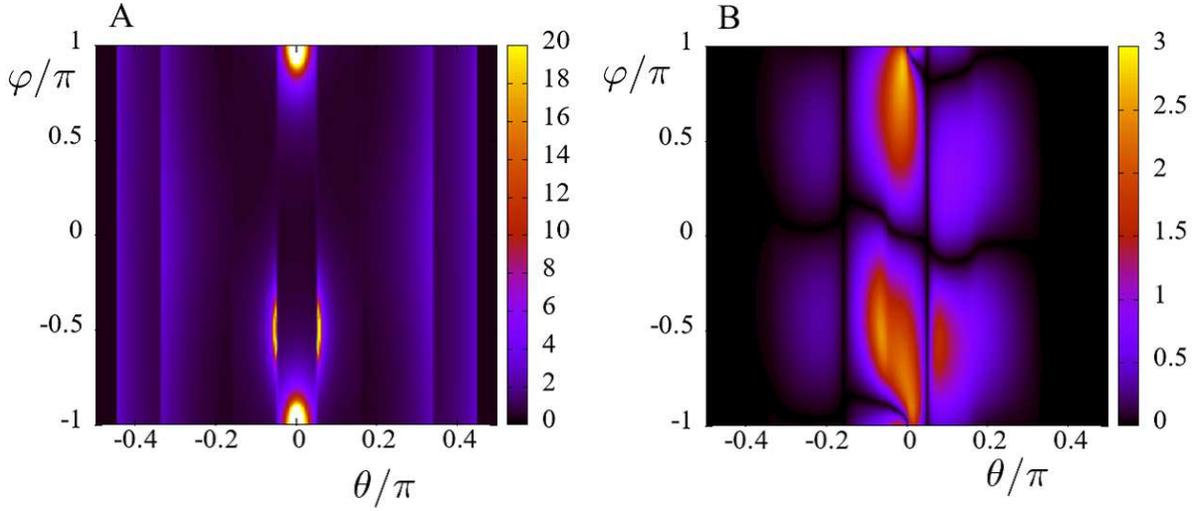}
			\end{center}
			\caption{
				The intensity plot of $S(0,\theta,\varphi)$ for 
				$(\alpha,\beta)=(-0.2\pi, 0.09\pi)$ in A.
				The angle resolved Josephson current 
$I(\theta,\varphi)$ = $ \mid I(\theta,\varphi,\alpha,\beta) 
				\mid $ is plotted in B. 
We choose $d \mid m_{z} \mid/v=1$, $\alpha=-0.2\pi$, $\beta=0.09\pi$  
and $m_{z}=0.5\mu$. }
\label{Fig8}
\end{figure}

 In order to understand the contribution of the zero energy Andreev bound states (ZEABS)  to Josephson current, 
		in Fig. \ref{Fig8}, 
		we plot $S(0,\theta,\varphi)$ and the magnitude of the 
		angle-resolved Josephson current 
		$\mid I(\theta, \varphi) \mid$ 
		with the same parameters used in Fig.\ref{Fig6}. 
		For the corresponding $\theta$ and $\varphi$ hosting 
		ZEABS, the resulting $S(E=0,\theta,\varphi)$ is enhanced.  
		Clearly, $\mid I(\theta, \varphi) \mid$ 
		is enhanced for $\theta$ and $\varphi$ when $S(0,\theta,\varphi)$ 
		shows the prominent peak structure. 
		Thus, the ZESAB and the angle resolved Josephson current has a correspondence. 
It is known from the study of $d$-wave superconductor junctions in the 
context of high T$_{c}$ cuprate, 
in the presence of the ZESABS, the 
Josephson current at low temperature  is mainly 
carried by ZESABS \cite{TKJosephson,TKJosephson2}. 
It can trigger non-monotonic temperature dependence of Josephson current 
observed in high Tc cuprate \cite{Testa}. The 
sophisticated $\theta$ and $\varphi$ dependence of 
$E_b(\theta, \varphi)$ in the present $d/FI/d$ junctions on TI 
shown in Figs. \ref{Fig6}, \ref{Fig7} and \ref{Fig8}, is due to the 
breaking of $\cal{P}$ and $\cal{T}$ symmetry, 
and generates the exotic current phase relation 
with simultaneous coexistence of 
$\sin\varphi$, $\sin2\varphi$ and $\cos\varphi$ terms. 

	\section{Conclusions and Discussions}
	In this paper, we have shown 
	a very large nonreciprocity of Josephson current 
	in a $d$-wave superconductor / Ferromagnetic insulator (FI)
	/ $d$-wave superconductor junction on  topological insulator.  
	We have found the large magnitude of quality factor 
	$Q$ which characterizes the diode effect 
	by tuning the crystal axis of both left and right 
	$d$-wave superconductors. \par
The magnitude of $Q$ becomes almost 0.4 at low temperatures 
and its sign is reversed by changing the 
direction of the magnetization in the FI. 
The physical origin of the large $Q$ stems from the 
exotic current-phase relations of the Josephson current 
due to the simultaneous existence of 
$\sin \varphi$, $\cos \varphi$ and $\sin 2\varphi$ 
component. 
The present situation is realized 
due to the strong asymmetry of the mirror inversion symmetry 
along the junction interface and the time reversal symmetry 
breaking by FI.  
		The strong temperature dependence of $Q$ 
		stems from the existence of the 
		low energy Andreev bound state appearing as  
		Majorana bound states (MBSs) at the interface.
		We have analyzed the Fourier components of Josephson current 
		and found that the $\cos \varphi$ 
		changes sign by the inversion of $m_{z}$. 
		These results can serve as a guide to design Josephson diode using 
		MBSs on the surface of TI. \par
	In this paper, we consider a two-dimensional (2D) junction. It is noted that the present diode effect does not exist in the 1D system. In this case, 
	only the contribution from $\theta$=0 remains 
	in the integral of $\theta$ in 
	eq. (\ref{currentphasesymmetry}). 
	Since we are considering even-parity superconductor,
	$\Gamma_{nL+}=\Gamma_{nL-}$ and $\Gamma_{nR+}=\Gamma_{nR-}$ 
	are satisfied 
	at $\theta=0$.  
	Then, $C(\theta=0)$ in eq. (\ref{BandC}) 
	becomes zero and the resulting $I(\varphi)$ does not have a $\cos \varphi$ 
	dependence. Then, we can not expect the present diode effect. 
	\par
	
	In the end, we mention the feasibility of the actual experiments. 
	The fabrication of the junction with 
	misorientation angles $\alpha \neq 0$ and $\beta \neq 0$ were realized in high $T_{c}$ cuprate to prove the $d$-wave nature of pairing 
	\cite{Tsuei1994,Tsuei2000}. 
	Also non-monotonic temperature dependence of the maximum Josephson current due to the enhanced $\sin 2\varphi$ component 
	was observed experimentally for $\alpha=-\beta \neq0$ \cite{Ilichev,Testa}. 
	On the other hand, Josephson current was observed in conventional $s$-wave superconductor junctions fabricated on the surface of TIs \cite{veldhorst12,williams12,Finck,Kurter2015}. 	It is noted that $4\pi$ periodicity due to the Kramers pair of MBS was reported \cite{Wiedenmann2016}. Furthermore, a high $T_{c}$ cuprate (Bi-2212) /TI junction was fabricated \cite{Zareapour2012}. 
	Based on these accumulated experimental works, the realization of the set-up in our proposal seems to be feasible, and our prediction can be tested in the near future. Finally, to pursue superconducting diode effect in the 
Josepshon junctions with topological superconductors is an interesting future 
issue \cite{Tanaka2012}. 
	
\begin{acknowledgments}
		We thank S. Tamura, T. Kokkeler and J.J. He for valuable discussions. 
		Y. T. was supported by 	Scientific Research (A) (KAKENHIGrant No. JP20H00131), 	and Scientific 	Research (B) (KAKENHIGrants No. JP18H01176 and No. JP20H01857). B. L was supported by National Natural Science Foundation of China (project 11904257) and the Natural Science Foundation of Tianjin (project 20JCQNJC01310). \par
	\end{acknowledgments}
	
		\section{Appndix 1}	
		In this section, we show eq. 	
		(\ref{currentinversion}). 
		We can show this relation analytically from 
		eqs. (\ref{currentphaseII}) and (\ref{formulaF}). 
		Since 
$\mid \Lambda_{dn}(\theta,\varphi) \mid^{2}
		=\mid \Lambda_{dn}(-\theta,\varphi) \mid^{2}$ is satisfied.  
		$I(\varphi)$ is expressed in 
		eq. \ref{currentphasesymmetry}. 
		Here $A(\theta)=A(-\theta)$, $B(\theta)=B(-\theta)$, and $C(\theta)=C(-\theta)$ are satisfied. 
		Since $\exp(-i \eta)$ changes into $\exp(i \eta)$ by the transformation of 
		$\theta$ to $-\theta$ or $m_{z}$ to $-m_{z}$, 
$\Lambda_{dn}(\theta,\varphi)
=\Lambda_{dn}(\theta,m_{z},\varphi)$ satisfies 
		\[
		\Lambda_{dn}\left(\theta,-m_{z},\varphi \right)
		=\Lambda_{dn}\left(-\theta,m_{z},-\varphi \right), \ 
		\Lambda_{dn}\left(-\theta,m_{z},\varphi \right)
		=\Lambda_{dn}^{*}\left(\theta,m_{z},\varphi \right)
		\]
		and 
		\begin{equation}
		\mid \Lambda_{dn}\left(\theta,-m_{z},\varphi \right)\mid^{2}=
		\mid \Lambda_{dn}\left(\theta,m_{z},-\varphi \right)\mid^{2}. 
		\end{equation}
		Also, 
		$\Lambda_{ne}=\Lambda_{ne}(\theta,m_{z})$ and $\Lambda_{no}=\Lambda_{no}(\theta,m_{z})$ satisfy 
		\[
		\Lambda_{ne}\left(\theta,m_{z} \right)=\Lambda_{ne}
		\left(\theta,-m_{z} \right), \ 
		\Lambda_{no}\left(\theta,m_{z} \right)=-\Lambda_{no}
		\left(\theta,-m_{z} \right). 
		\]
		Then, $A(\theta)=A(\theta,m_{z})$, $B(\theta)=B(\theta,m_{z})$,
 and $C(\theta)=C(\theta,m_{z})$ 
		satisfy
\[
A\left(\theta,m_{z}\right)=A\left(\theta, -m_{z}\right), 
\ 
B\left(\theta,m_{z}\right)=B\left(\theta, -m_{z}\right),
\ C\left(\theta,m_{z} \right)=-C\left(\theta,-m_{z} \right). 
\]
As a result, we can derive eq. (\ref{currentinversion}). \par			
		
		\section{Appendix 2}
		We can show eq. (\ref{Qinversion})  as follows. 
		Since $\Gamma_{nL\pm}$ and $\Gamma_{nR\pm}$ change into
		$\Gamma_{nL\mp}$ and $\Gamma_{nR\mp}$
		by the transformation $(\alpha,\beta)$ to $(-\alpha,-\beta)$,  
		$\Lambda_{dn}(\theta,\varphi)=\Lambda_{dn}(\theta,\varphi,\alpha,\beta)$, 
		$\Lambda_{ne}=\Lambda_{ne}(\theta,\alpha,\beta)$, and 
		$\Lambda_{no}=\Lambda_{no}(\theta,\alpha,\beta)$ in eqs. 
		(\ref{Lambdadn}), (\ref{Lambdane}), and (\ref{Lambdano}) satisfy
		\begin{equation}
		\Lambda_{dn}\left(\theta,\varphi,-\alpha,-\beta \right)
		=\Lambda_{dn}\left(\theta,-\varphi,\alpha,\beta \right), 
		\end{equation}
		and
		\begin{equation}
		\Lambda_{ne}\left(\theta,-\alpha,-\beta \right)
		=\Lambda_{ne}\left(\theta,\alpha,\beta \right), \ \ 
		\Lambda_{no}\left(\theta,-\alpha,-\beta \right)
		=\Lambda_{no}\left(\theta,\alpha,\beta \right). 
		\end{equation}
		If we write $\alpha$, $\beta$ dependence of $I(\varphi)$ explicitly, 
		\begin{equation}
		I\left(\varphi, \alpha,\beta \right)
		= - I\left(-\varphi,-\alpha,-\beta \right)
		\end{equation}
		is obtained and  
		it leads to eq. (\ref{Qinversion}). 
	
	\section{Appendix3}
	In this Appendix, we explain why simple $d$-wave superconductor junctions by 
	cuprate without TI does not show any diode effet. 
	In $d$-wave / ferromagnet insulator  /$d$-wave superconductor junction,
	there is no diode effect since $\cos \varphi$ term is not generated as shown in Ref. \cite{Tanaka2000} if the spin-orbit coupling is absent.
	We can prove why 
	d/FI/d junction without TI can not hold the diode effect.
	The Hamiltonian in $d$-wave junctions realized in cuprates
	is given by
	\begin{equation}
	\left[
	\begin{array}{cccc}
	-\frac{\hbar ^{2}\left( \partial _{x}^{2}+\partial _{y}^{2}\right) }{2m}-\mu
	+m_{z} & 0 & 0 & \Delta \left( k_{x},k_{y}\right) e^{i\varphi } \\
	0 & -\frac{\hbar ^{2}\left( \partial _{x}^{2}+\partial _{y}^{2}\right) }{2m}%
	-\mu -m_{z} & -\Delta \left( k_{x},k_{y}\right) e^{i\varphi } & 0 \\
	0 & -\Delta \left( k_{x},k_{y}\right) e^{-i\varphi } & \frac{\hbar
		^{2}\left( \partial _{x}^{2}+\partial _{y}^{2}\right) }{2m}+\mu -m_{z} & 0
	\\
	\Delta \left( k_{x},k_{y}\right) e^{-i\varphi } & 0 & 0 & \frac{\hbar
		^{2}\left( \partial _{x}^{2}+\partial _{y}^{2}\right) }{2m}+\mu +m_{z}%
	\end{array}%
	\right] ,
	\end{equation}%
	with%
	\begin{equation}
	\Delta \left( k_{x},k_{y}\right) =\left( \hat{k}_{x}^{2}-\hat{k}%
	_{y}^{2}\right) \cos 2\alpha -2\hat{k}_{x}\hat{k}_{y}\sin 2\alpha .
	\end{equation}%
	The time reversal symmetry is%
	\begin{equation}
	T=\left[
	\begin{array}{cc}
	is_{y}K & 0 \\
	0 & is_{y}K%
	\end{array}%
	\right] ,
	\end{equation}%
	and another relevant operator is the $C_{2y}$ given by
	\begin{equation}
	C_{2y}=\left[
	\begin{array}{cc}
	e^{-i\frac{\pi }{2}s_{y}} & 0 \\
	0 & e^{-i\frac{\pi }{2}s_{y}}%
	\end{array}%
	\right] =\left[
	\begin{array}{cc}
	-is_{y} & 0 \\
	0 & -is_{y}%
	\end{array}%
	\right] .
	\end{equation}%
	Let us define a combined operator $\tilde{T}$
	\begin{equation}
	\tilde{T}=TC_{2y}=\left[
	\begin{array}{cc}
	-s_{0}K & 0 \\
	0 & -s_{0}K%
	\end{array}%
	\right] ,
	\end{equation}%
	we can obtain%
	\begin{equation}
	\tilde{T}H\left( \varphi \right) \tilde{T}^{-1}=H\left( -\varphi \right).
	\end{equation}%
	It implies that the energy spectrum has symmetry
	\begin{equation}
	E_{n}\left( \varphi \right) =E_{n}\left( -\varphi \right) .
	\end{equation}%
	The energy of the junction is an even function of the phase difference
	$\varphi$. Thus, the Josephson current is an odd function of $\varphi$
	according to
	\begin{equation}
	I\left( \varphi \right) =\frac{2e}{\hbar }\frac{\partial F}{\partial \varphi
	}=\frac{2e}{\hbar }\frac{\partial }{\partial \varphi }\left(
	\sum\nolimits_{n}E_{n}\left( \varphi \right) f_{n}\right),
	\end{equation}%
	\begin{equation}
	I\left( -\varphi \right) =\frac{2e}{\hbar }
	\frac{\partial F}{\partial \varphi }
	=\frac{2e}{\hbar }\frac{\partial }{\partial \varphi }\left(
	\sum\nolimits_{n}E_{n}\left( -\varphi \right) f_{n}\right) =-I\left( \varphi
	\right)
	\end{equation}%
with Fermi distribution function $f_{n}$. 
	Then, we can get%
	\begin{equation}
	I\left( -\varphi \right) =-I\left( \varphi \right) ,
	\end{equation}%
	and%
	\begin{equation}
	I\left( 0\right) =0.
	\end{equation}%
	If $I\left( 0\right) =0$, the Josephson current can not hold
	$\cos \varphi$ term which is required by diode effect.
	We can therefore conclude that the
	similar d/FI/d junction without TI can not harbor diode effect due
	to the combined $\tilde{T}$ symmetry. Instead, if there is spin-orbit
	couplings, $\tilde{T}H\left( \varphi \right) \tilde{T}^{-1}$ will no longer
	equal to $H\left( -\varphi \right)$, then we can expect the $\cos \varphi$
	term and diode effect.
	The presence of spin-orbit coupling is essential for the diode effect. \par
	On the other hand, the surface state of a topological insulator
	has a strong spin-orbit coupling which generates spin-momentum locking.
	To enhance $\cos\varphi$ term, it is promising to consider
	$d/FI/d$ junction on the surface of TI.

\bibliography{TopologicalSC}

\begin{thebibliography}{58}%
\makeatletter
\providecommand \@ifxundefined [1]{%
 \@ifx{#1\undefined}
}%
\providecommand \@ifnum [1]{%
 \ifnum #1\expandafter \@firstoftwo
 \else \expandafter \@secondoftwo
 \fi
}%
\providecommand \@ifx [1]{%
 \ifx #1\expandafter \@firstoftwo
 \else \expandafter \@secondoftwo
 \fi
}%
\providecommand \natexlab [1]{#1}%
\providecommand \enquote  [1]{``#1''}%
\providecommand \bibnamefont  [1]{#1}%
\providecommand \bibfnamefont [1]{#1}%
\providecommand \citenamefont [1]{#1}%
\providecommand \href@noop [0]{\@secondoftwo}%
\providecommand \href [0]{\begingroup \@sanitize@url \@href}%
\providecommand \@href[1]{\@@startlink{#1}\@@href}%
\providecommand \@@href[1]{\endgroup#1\@@endlink}%
\providecommand \@sanitize@url [0]{\catcode `\\12\catcode `\$12\catcode
  `\&12\catcode `\#12\catcode `\^12\catcode `\_12\catcode `\%12\relax}%
\providecommand \@@startlink[1]{}%
\providecommand \@@endlink[0]{}%
\providecommand \url  [0]{\begingroup\@sanitize@url \@url }%
\providecommand \@url [1]{\endgroup\@href {#1}{\urlprefix }}%
\providecommand \urlprefix  [0]{URL }%
\providecommand \Eprint [0]{\href }%
\providecommand \doibase [0]{http://dx.doi.org/}%
\providecommand \selectlanguage [0]{\@gobble}%
\providecommand \bibinfo  [0]{\@secondoftwo}%
\providecommand \bibfield  [0]{\@secondoftwo}%
\providecommand \translation [1]{[#1]}%
\providecommand \BibitemOpen [0]{}%
\providecommand \bibitemStop [0]{}%
\providecommand \bibitemNoStop [0]{.\EOS\space}%
\providecommand \EOS [0]{\spacefactor3000\relax}%
\providecommand \BibitemShut  [1]{\csname bibitem#1\endcsname}%
\let\auto@bib@innerbib\@empty
\bibitem [{\citenamefont {Tokura}\ and\ \citenamefont
  {Nagaosa}(2018)}]{TokuraNagaosa}%
  \BibitemOpen
  \bibfield  {author} {\bibinfo {author} {\bibfnamefont {Y.}~\bibnamefont
  {Tokura}}\ and\ \bibinfo {author} {\bibfnamefont {N.}~\bibnamefont
  {Nagaosa}},\ }\href {\doibase 10.1038/s41467-018-05759-4} {\bibfield
  {journal} {\bibinfo  {journal} {Nat. Commun.}\ }\textbf {\bibinfo {volume}
  {9}},\ \bibinfo {pages} {3740} (\bibinfo {year} {2018})}\BibitemShut
  {NoStop}%
\bibitem [{\citenamefont {Rikken}\ \emph {et~al.}(2001)\citenamefont {Rikken},
  \citenamefont {F\"olling},\ and\ \citenamefont {Wyder}}]{Rikken1}%
  \BibitemOpen
  \bibfield  {author} {\bibinfo {author} {\bibfnamefont {G.~L. J.~A.}\
  \bibnamefont {Rikken}}, \bibinfo {author} {\bibfnamefont {J.}~\bibnamefont
  {F\"olling}}, \ and\ \bibinfo {author} {\bibfnamefont {P.}~\bibnamefont
  {Wyder}},\ }\href {\doibase 10.1103/PhysRevLett.87.236602} {\bibfield
  {journal} {\bibinfo  {journal} {Phys. Rev. Lett.}\ }\textbf {\bibinfo
  {volume} {87}},\ \bibinfo {pages} {236602} (\bibinfo {year}
  {2001})}\BibitemShut {NoStop}%
\bibitem [{\citenamefont {Krsti\'{c}}\ \emph {et~al.}(2002)\citenamefont
  {Krsti\'{c}}, \citenamefont {Roth}, \citenamefont {Burghard}, \citenamefont
  {Kern},\ and\ \citenamefont {Rikken}}]{Rikken2}%
  \BibitemOpen
  \bibfield  {author} {\bibinfo {author} {\bibfnamefont {V.}~\bibnamefont
  {Krsti\'{c}}}, \bibinfo {author} {\bibfnamefont {S.}~\bibnamefont {Roth}},
  \bibinfo {author} {\bibfnamefont {M.}~\bibnamefont {Burghard}}, \bibinfo
  {author} {\bibfnamefont {K.}~\bibnamefont {Kern}}, \ and\ \bibinfo {author}
  {\bibfnamefont {G.~L. J.~A.}\ \bibnamefont {Rikken}},\ }\href {\doibase
  10.1063/1.1523895} {\bibfield  {journal} {\bibinfo  {journal} {J. Chem.
  Phys.}\ }\textbf {\bibinfo {volume} {117}},\ \bibinfo {pages} {11315}
  (\bibinfo {year} {2002})}\BibitemShut {NoStop}%
\bibitem [{\citenamefont {Pop}\ \emph {et~al.}(2014)\citenamefont {Pop},
  \citenamefont {Auban-Senzier}, \citenamefont {Canadell}, \citenamefont
  {Rikken},\ and\ \citenamefont {Avarvari}}]{Rikken3}%
  \BibitemOpen
  \bibfield  {author} {\bibinfo {author} {\bibfnamefont {F.}~\bibnamefont
  {Pop}}, \bibinfo {author} {\bibfnamefont {P.}~\bibnamefont {Auban-Senzier}},
  \bibinfo {author} {\bibfnamefont {E.}~\bibnamefont {Canadell}}, \bibinfo
  {author} {\bibfnamefont {G.~L. J.~A.}\ \bibnamefont {Rikken}}, \ and\
  \bibinfo {author} {\bibfnamefont {N.}~\bibnamefont {Avarvari}},\ }\href
  {\doibase 10.1038/ncomms4757} {\bibfield  {journal} {\bibinfo  {journal}
  {Nat. Commun.}\ }\textbf {\bibinfo {volume} {5}},\ \bibinfo {pages} {3757}
  (\bibinfo {year} {2014})}\BibitemShut {NoStop}%
\bibitem [{\citenamefont {Rikken}\ and\ \citenamefont {Wyder}(2005)}]{Rikken4}%
  \BibitemOpen
  \bibfield  {author} {\bibinfo {author} {\bibfnamefont {G.~L. J.~A.}\
  \bibnamefont {Rikken}}\ and\ \bibinfo {author} {\bibfnamefont
  {P.}~\bibnamefont {Wyder}},\ }\href {\doibase 10.1103/PhysRevLett.94.016601}
  {\bibfield  {journal} {\bibinfo  {journal} {Phys. Rev. Lett.}\ }\textbf
  {\bibinfo {volume} {94}},\ \bibinfo {pages} {016601} (\bibinfo {year}
  {2005})}\BibitemShut {NoStop}%
\bibitem [{\citenamefont {Ideue}\ \emph {et~al.}(2017)\citenamefont {Ideue},
  \citenamefont {Hamamoto}, \citenamefont {Koshikawa}, \citenamefont {Ezawa},
  \citenamefont {Shimizu}, \citenamefont {Kaneko}, \citenamefont {Tokura},
  \citenamefont {Nagaosa},\ and\ \citenamefont {Iwasa}}]{IwasaNatPhys}%
  \BibitemOpen
  \bibfield  {author} {\bibinfo {author} {\bibfnamefont {T.}~\bibnamefont
  {Ideue}}, \bibinfo {author} {\bibfnamefont {K.}~\bibnamefont {Hamamoto}},
  \bibinfo {author} {\bibfnamefont {S.}~\bibnamefont {Koshikawa}}, \bibinfo
  {author} {\bibfnamefont {M.}~\bibnamefont {Ezawa}}, \bibinfo {author}
  {\bibfnamefont {S.}~\bibnamefont {Shimizu}}, \bibinfo {author} {\bibfnamefont
  {Y.}~\bibnamefont {Kaneko}}, \bibinfo {author} {\bibfnamefont
  {Y.}~\bibnamefont {Tokura}}, \bibinfo {author} {\bibfnamefont
  {N.}~\bibnamefont {Nagaosa}}, \ and\ \bibinfo {author} {\bibfnamefont
  {Y.}~\bibnamefont {Iwasa}},\ }\href {\doibase 10.1038/nphys4056} {\bibfield
  {journal} {\bibinfo  {journal} {Nat. Phys.}\ }\textbf {\bibinfo {volume}
  {13}},\ \bibinfo {pages} {578} (\bibinfo {year} {2017})}\BibitemShut
  {NoStop}%
\bibitem [{\citenamefont {Hoshino}\ \emph {et~al.}(2018)\citenamefont
  {Hoshino}, \citenamefont {Wakatsuki}, \citenamefont {Hamamoto},\ and\
  \citenamefont {Nagaosa}}]{Hoshino}%
  \BibitemOpen
  \bibfield  {author} {\bibinfo {author} {\bibfnamefont {S.}~\bibnamefont
  {Hoshino}}, \bibinfo {author} {\bibfnamefont {R.}~\bibnamefont {Wakatsuki}},
  \bibinfo {author} {\bibfnamefont {K.}~\bibnamefont {Hamamoto}}, \ and\
  \bibinfo {author} {\bibfnamefont {N.}~\bibnamefont {Nagaosa}},\ }\href
  {\doibase 10.1103/PhysRevB.98.054510} {\bibfield  {journal} {\bibinfo
  {journal} {Phys. Rev. B}\ }\textbf {\bibinfo {volume} {98}},\ \bibinfo
  {pages} {054510} (\bibinfo {year} {2018})}\BibitemShut {NoStop}%
\bibitem [{\citenamefont {Wakatsuki}\ \emph {et~al.}(2017)\citenamefont
  {Wakatsuki}, \citenamefont {Saito}, \citenamefont {Hoshino}, \citenamefont
  {Itahashi}, \citenamefont {Ideue}, \citenamefont {Ezawa}, \citenamefont
  {Iwasa},\ and\ \citenamefont {Nagaosa}}]{WakatsukiSciAdv}%
  \BibitemOpen
  \bibfield  {author} {\bibinfo {author} {\bibfnamefont {R.}~\bibnamefont
  {Wakatsuki}}, \bibinfo {author} {\bibfnamefont {Y.}~\bibnamefont {Saito}},
  \bibinfo {author} {\bibfnamefont {S.}~\bibnamefont {Hoshino}}, \bibinfo
  {author} {\bibfnamefont {Y.~M.}\ \bibnamefont {Itahashi}}, \bibinfo {author}
  {\bibfnamefont {T.}~\bibnamefont {Ideue}}, \bibinfo {author} {\bibfnamefont
  {M.}~\bibnamefont {Ezawa}}, \bibinfo {author} {\bibfnamefont
  {Y.}~\bibnamefont {Iwasa}}, \ and\ \bibinfo {author} {\bibfnamefont
  {N.}~\bibnamefont {Nagaosa}},\ }\href {\doibase 10.1126/sciadv.1602390}
  {\bibfield  {journal} {\bibinfo  {journal} {Sci. Adv.}\ }\textbf {\bibinfo
  {volume} {3}},\ \bibinfo {pages} {e1602390} (\bibinfo {year}
  {2017})}\BibitemShut {NoStop}%
\bibitem [{\citenamefont {Qin}\ \emph {et~al.}(2017)\citenamefont {Qin},
  \citenamefont {Shi}, \citenamefont {Ideue}, \citenamefont {Yoshida},
  \citenamefont {Zak}, \citenamefont {Tenne}, \citenamefont {Kikitsu},
  \citenamefont {Inoue}, \citenamefont {Hashizume},\ and\ \citenamefont
  {Iwasa}}]{IwasaNatC}%
  \BibitemOpen
  \bibfield  {author} {\bibinfo {author} {\bibfnamefont {F.}~\bibnamefont
  {Qin}}, \bibinfo {author} {\bibfnamefont {W.}~\bibnamefont {Shi}}, \bibinfo
  {author} {\bibfnamefont {T.}~\bibnamefont {Ideue}}, \bibinfo {author}
  {\bibfnamefont {M.}~\bibnamefont {Yoshida}}, \bibinfo {author} {\bibfnamefont
  {A.}~\bibnamefont {Zak}}, \bibinfo {author} {\bibfnamefont {R.}~\bibnamefont
  {Tenne}}, \bibinfo {author} {\bibfnamefont {T.}~\bibnamefont {Kikitsu}},
  \bibinfo {author} {\bibfnamefont {D.}~\bibnamefont {Inoue}}, \bibinfo
  {author} {\bibfnamefont {D.}~\bibnamefont {Hashizume}}, \ and\ \bibinfo
  {author} {\bibfnamefont {Y.}~\bibnamefont {Iwasa}},\ }\href {\doibase
  10.1038/ncomms14465} {\bibfield  {journal} {\bibinfo  {journal} {Nat.
  Commun.}\ }\textbf {\bibinfo {volume} {8}},\ \bibinfo {pages} {14465}
  (\bibinfo {year} {2017})}\BibitemShut {NoStop}%
\bibitem [{\citenamefont {Itahashi}\ \emph {et~al.}(2020)\citenamefont
  {Itahashi}, \citenamefont {Ideue}, \citenamefont {Saito}, \citenamefont
  {Shimizu}, \citenamefont {Ouchi}, \citenamefont {Nojima},\ and\ \citenamefont
  {Iwasa}}]{Itahashi}%
  \BibitemOpen
  \bibfield  {author} {\bibinfo {author} {\bibfnamefont {Y.~M.}\ \bibnamefont
  {Itahashi}}, \bibinfo {author} {\bibfnamefont {T.}~\bibnamefont {Ideue}},
  \bibinfo {author} {\bibfnamefont {Y.}~\bibnamefont {Saito}}, \bibinfo
  {author} {\bibfnamefont {S.}~\bibnamefont {Shimizu}}, \bibinfo {author}
  {\bibfnamefont {T.}~\bibnamefont {Ouchi}}, \bibinfo {author} {\bibfnamefont
  {T.}~\bibnamefont {Nojima}}, \ and\ \bibinfo {author} {\bibfnamefont
  {Y.}~\bibnamefont {Iwasa}},\ }\href {\doibase 10.1126/sciadv.aay9120}
  {\bibfield  {journal} {\bibinfo  {journal} {Sci. Adv.}\ }\textbf {\bibinfo
  {volume} {6}},\ \bibinfo {pages} {eaay9120} (\bibinfo {year}
  {2020})}\BibitemShut {NoStop}%
\bibitem [{\citenamefont {Ando}\ \emph {et~al.}(2020)\citenamefont {Ando},
  \citenamefont {Miyasaka}, \citenamefont {Li}, \citenamefont {Ishizuka},
  \citenamefont {Arakawa}, \citenamefont {Shiota}, \citenamefont {Moriyama},
  \citenamefont {Yanase},\ and\ \citenamefont {Ono}}]{OnoNat}%
  \BibitemOpen
  \bibfield  {author} {\bibinfo {author} {\bibfnamefont {F.}~\bibnamefont
  {Ando}}, \bibinfo {author} {\bibfnamefont {Y.}~\bibnamefont {Miyasaka}},
  \bibinfo {author} {\bibfnamefont {T.}~\bibnamefont {Li}}, \bibinfo {author}
  {\bibfnamefont {J.}~\bibnamefont {Ishizuka}}, \bibinfo {author}
  {\bibfnamefont {T.}~\bibnamefont {Arakawa}}, \bibinfo {author} {\bibfnamefont
  {Y.}~\bibnamefont {Shiota}}, \bibinfo {author} {\bibfnamefont
  {T.}~\bibnamefont {Moriyama}}, \bibinfo {author} {\bibfnamefont
  {Y.}~\bibnamefont {Yanase}}, \ and\ \bibinfo {author} {\bibfnamefont
  {T.}~\bibnamefont {Ono}},\ }\href {\doibase 10.1038/s41586-020-2590-4}
  {\bibfield  {journal} {\bibinfo  {journal} {Nature}\ }\textbf {\bibinfo
  {volume} {584}},\ \bibinfo {pages} {373} (\bibinfo {year}
  {2020})}\BibitemShut {NoStop}%
\bibitem [{\citenamefont {Pal}\ \emph {et~al.}(2022)\citenamefont {Pal},
  \citenamefont {Chakraborty}, \citenamefont {Sivakumar}, \citenamefont
  {Davydova}, \citenamefont {Gopi}, \citenamefont {Pandeya}, \citenamefont
  {Krieger}, \citenamefont {Zhang}, \citenamefont {Date}, \citenamefont {Ju},
  \citenamefont {Yuan}, \citenamefont {Schroter}, \citenamefont {Fu},\ and\
  \citenamefont {Parkin}}]{Pal2022}%
  \BibitemOpen
  \bibfield  {author} {\bibinfo {author} {\bibfnamefont {B.}~\bibnamefont
  {Pal}}, \bibinfo {author} {\bibfnamefont {A.}~\bibnamefont {Chakraborty}},
  \bibinfo {author} {\bibfnamefont {P.~K.}\ \bibnamefont {Sivakumar}}, \bibinfo
  {author} {\bibfnamefont {M.}~\bibnamefont {Davydova}}, \bibinfo {author}
  {\bibfnamefont {A.~K.}\ \bibnamefont {Gopi}}, \bibinfo {author}
  {\bibfnamefont {A.~K.}\ \bibnamefont {Pandeya}}, \bibinfo {author}
  {\bibfnamefont {J.~A.}\ \bibnamefont {Krieger}}, \bibinfo {author}
  {\bibfnamefont {Y.}~\bibnamefont {Zhang}}, \bibinfo {author} {\bibfnamefont
  {M.}~\bibnamefont {Date}}, \bibinfo {author} {\bibfnamefont {S.}~\bibnamefont
  {Ju}}, \bibinfo {author} {\bibfnamefont {N.}~\bibnamefont {Yuan}}, \bibinfo
  {author} {\bibfnamefont {N.~B.~M.}\ \bibnamefont {Schroter}}, \bibinfo
  {author} {\bibfnamefont {L.}~\bibnamefont {Fu}}, \ and\ \bibinfo {author}
  {\bibfnamefont {S.~S.~P.}\ \bibnamefont {Parkin}},\ }\href {\doibase
  10.1038/s41567-022-01699-5} {\bibfield  {journal} {\bibinfo  {journal}
  {Nature Physics}\ }\textbf {\bibinfo {volume} {18}},\ \bibinfo {pages} {1228}
  (\bibinfo {year} {2022})}\BibitemShut {NoStop}%
\bibitem [{\citenamefont {Narita}\ \emph {et~al.}(2022)\citenamefont {Narita},
  \citenamefont {Ishizuka}, \citenamefont {Kawarazaki}, \citenamefont {Kan},
  \citenamefont {Shiota}, \citenamefont {Moriyama}, \citenamefont {Shimakawa},
  \citenamefont {Ognev}, \citenamefont {Samardak}, \citenamefont {Yanase},\
  and\ \citenamefont {Ono}}]{Narita2022}%
  \BibitemOpen
  \bibfield  {author} {\bibinfo {author} {\bibfnamefont {H.}~\bibnamefont
  {Narita}}, \bibinfo {author} {\bibfnamefont {J.}~\bibnamefont {Ishizuka}},
  \bibinfo {author} {\bibfnamefont {R.}~\bibnamefont {Kawarazaki}}, \bibinfo
  {author} {\bibfnamefont {D.}~\bibnamefont {Kan}}, \bibinfo {author}
  {\bibfnamefont {Y.}~\bibnamefont {Shiota}}, \bibinfo {author} {\bibfnamefont
  {T.}~\bibnamefont {Moriyama}}, \bibinfo {author} {\bibfnamefont
  {Y.}~\bibnamefont {Shimakawa}}, \bibinfo {author} {\bibfnamefont {A.~V.}\
  \bibnamefont {Ognev}}, \bibinfo {author} {\bibfnamefont {A.~S.}\ \bibnamefont
  {Samardak}}, \bibinfo {author} {\bibfnamefont {Y.}~\bibnamefont {Yanase}}, \
  and\ \bibinfo {author} {\bibfnamefont {T.}~\bibnamefont {Ono}},\ }\href
  {\doibase 10.1038/s41565-022-01159-4} {\bibfield  {journal} {\bibinfo
  {journal} {Nature Nanotechnology}\ }\textbf {\bibinfo {volume} {17}},\
  \bibinfo {pages} {823} (\bibinfo {year} {2022})}\BibitemShut {NoStop}%
\bibitem [{\citenamefont {Jeon}\ \emph {et~al.}(2022)\citenamefont {Jeon},
  \citenamefont {Kim}, \citenamefont {Yoon}, \citenamefont {Jeon},
  \citenamefont {Han}, \citenamefont {Cottet}, \citenamefont {Kontos},\ and\
  \citenamefont {Parkin}}]{Jeon2022}%
  \BibitemOpen
  \bibfield  {author} {\bibinfo {author} {\bibfnamefont {K.-R.}\ \bibnamefont
  {Jeon}}, \bibinfo {author} {\bibfnamefont {J.-K.}\ \bibnamefont {Kim}},
  \bibinfo {author} {\bibfnamefont {J.}~\bibnamefont {Yoon}}, \bibinfo {author}
  {\bibfnamefont {J.-C.}\ \bibnamefont {Jeon}}, \bibinfo {author}
  {\bibfnamefont {H.}~\bibnamefont {Han}}, \bibinfo {author} {\bibfnamefont
  {A.}~\bibnamefont {Cottet}}, \bibinfo {author} {\bibfnamefont
  {T.}~\bibnamefont {Kontos}}, \ and\ \bibinfo {author} {\bibfnamefont
  {S.~S.~P.}\ \bibnamefont {Parkin}},\ }\href {\doibase
  10.1038/s41563-022-01300-7} {\bibfield  {journal} {\bibinfo  {journal}
  {Nature Materials}\ }\textbf {\bibinfo {volume} {21}},\ \bibinfo {pages}
  {1008} (\bibinfo {year} {2022})}\BibitemShut {NoStop}%
\bibitem [{\citenamefont {Bauriedl}\ \emph {et~al.}(2022)\citenamefont
  {Bauriedl}, \citenamefont {B"{a}uml}, \citenamefont {Fuchs}, \citenamefont
  {Baumgartner}, \citenamefont {Paulik}, \citenamefont {Bauer}, \citenamefont
  {Lin}, \citenamefont {Lupton}, \citenamefont {Taniguchi}, \citenamefont
  {Watanabe}, \citenamefont {Strunk},\ and\ \citenamefont
  {Paradiso}}]{Bauriedl2022}%
  \BibitemOpen
  \bibfield  {author} {\bibinfo {author} {\bibfnamefont {L.}~\bibnamefont
  {Bauriedl}}, \bibinfo {author} {\bibfnamefont {C.}~\bibnamefont {B"{a}uml}},
  \bibinfo {author} {\bibfnamefont {L.}~\bibnamefont {Fuchs}}, \bibinfo
  {author} {\bibfnamefont {C.}~\bibnamefont {Baumgartner}}, \bibinfo {author}
  {\bibfnamefont {N.}~\bibnamefont {Paulik}}, \bibinfo {author} {\bibfnamefont
  {J.~M.}\ \bibnamefont {Bauer}}, \bibinfo {author} {\bibfnamefont {K.-Q.}\
  \bibnamefont {Lin}}, \bibinfo {author} {\bibfnamefont {J.~M.}\ \bibnamefont
  {Lupton}}, \bibinfo {author} {\bibfnamefont {T.}~\bibnamefont {Taniguchi}},
  \bibinfo {author} {\bibfnamefont {K.}~\bibnamefont {Watanabe}}, \bibinfo
  {author} {\bibfnamefont {C.}~\bibnamefont {Strunk}}, \ and\ \bibinfo {author}
  {\bibfnamefont {N.}~\bibnamefont {Paradiso}},\ }\href {\doibase
  10.1038/s41467-022-31954-5} {\bibfield  {journal} {\bibinfo  {journal}
  {Nature Communications}\ }\textbf {\bibinfo {volume} {13}},\ \bibinfo {pages}
  {4266} (\bibinfo {year} {2022})}\BibitemShut {NoStop}%
\bibitem [{\citenamefont {He}\ \emph {et~al.}(2022)\citenamefont {He},
  \citenamefont {Tanaka},\ and\ \citenamefont {Nagaosa}}]{James2022}%
  \BibitemOpen
  \bibfield  {author} {\bibinfo {author} {\bibfnamefont {J.~J.}\ \bibnamefont
  {He}}, \bibinfo {author} {\bibfnamefont {Y.}~\bibnamefont {Tanaka}}, \ and\
  \bibinfo {author} {\bibfnamefont {N.}~\bibnamefont {Nagaosa}},\ }\href
  {\doibase 10.1088/1367-2630/ac6766} {\bibfield  {journal} {\bibinfo
  {journal} {New J. Phys.}\ }\textbf {\bibinfo {volume} {24}},\ \bibinfo
  {pages} {053014} (\bibinfo {year} {2022})}\BibitemShut {NoStop}%
\bibitem [{\citenamefont {Daido}\ \emph {et~al.}(2022)\citenamefont {Daido},
  \citenamefont {Ikeda},\ and\ \citenamefont {Yanase}}]{Daido}%
  \BibitemOpen
  \bibfield  {author} {\bibinfo {author} {\bibfnamefont {A.}~\bibnamefont
  {Daido}}, \bibinfo {author} {\bibfnamefont {Y.}~\bibnamefont {Ikeda}}, \ and\
  \bibinfo {author} {\bibfnamefont {Y.}~\bibnamefont {Yanase}},\ }\href
  {\doibase 10.1103/PhysRevLett.128.037001} {\bibfield  {journal} {\bibinfo
  {journal} {Phys. Rev. Lett.}\ }\textbf {\bibinfo {volume} {128}},\ \bibinfo
  {pages} {037001} (\bibinfo {year} {2022})}\BibitemShut {NoStop}%
\bibitem [{\citenamefont {Yuan}\ and\ \citenamefont {Fu}(2022)}]{Fu}%
  \BibitemOpen
  \bibfield  {author} {\bibinfo {author} {\bibfnamefont {N.~F.~Q.}\
  \bibnamefont {Yuan}}\ and\ \bibinfo {author} {\bibfnamefont {L.}~\bibnamefont
  {Fu}},\ }\href {\doibase 10.1073/pnas.2119548119} {\bibfield  {journal}
  {\bibinfo  {journal} {Proc. Nat. Acad. of Sci.}\ }\textbf {\bibinfo {volume}
  {119}},\ \bibinfo {pages} {e2119548119} (\bibinfo {year} {2022})}\BibitemShut
  {NoStop}%
\bibitem [{\citenamefont {Ili\ifmmode~\acute{c}\else \'{c}\fi{}}\ and\
  \citenamefont {Bergeret}(2022)}]{Bergeret}%
  \BibitemOpen
  \bibfield  {author} {\bibinfo {author} {\bibfnamefont {S.}~\bibnamefont
  {Ili\ifmmode~\acute{c}\else \'{c}\fi{}}}\ and\ \bibinfo {author}
  {\bibfnamefont {F.~S.}\ \bibnamefont {Bergeret}},\ }\href {\doibase
  10.1103/PhysRevLett.128.177001} {\bibfield  {journal} {\bibinfo  {journal}
  {Phys. Rev. Lett.}\ }\textbf {\bibinfo {volume} {128}},\ \bibinfo {pages}
  {177001} (\bibinfo {year} {2022})}\BibitemShut {NoStop}%
\bibitem [{\citenamefont {Karabassov}\ \emph {et~al.}(2022)\citenamefont
  {Karabassov}, \citenamefont {Bobkova}, \citenamefont {Golubov},\ and\
  \citenamefont {Vasenko}}]{Bovkova}%
  \BibitemOpen
  \bibfield  {author} {\bibinfo {author} {\bibfnamefont {T.}~\bibnamefont
  {Karabassov}}, \bibinfo {author} {\bibfnamefont {I.~V.}\ \bibnamefont
  {Bobkova}}, \bibinfo {author} {\bibfnamefont {A.~A.}\ \bibnamefont
  {Golubov}}, \ and\ \bibinfo {author} {\bibfnamefont {A.~S.}\ \bibnamefont
  {Vasenko}},\ }\href {http://arxiv.org/abs/2203.15608} {\bibfield  {journal}
  {\bibinfo  {journal} {arXiv:2203.15608}\ } (\bibinfo {year}
  {2022})}\BibitemShut {NoStop}%
\bibitem [{\citenamefont {Souto}\ \emph {et~al.}(2022)\citenamefont {Souto},
  \citenamefont {Leijnse},\ and\ \citenamefont {Schrade}}]{Schrade}%
  \BibitemOpen
  \bibfield  {author} {\bibinfo {author} {\bibfnamefont {R.~S.}\ \bibnamefont
  {Souto}}, \bibinfo {author} {\bibfnamefont {M.}~\bibnamefont {Leijnse}}, \
  and\ \bibinfo {author} {\bibfnamefont {C.}~\bibnamefont {Schrade}},\ }\href
  {http://arxiv.org/abs/2205.04469} {\bibfield  {journal} {\bibinfo  {journal}
  {arXiv:2205.04469}\ } (\bibinfo {year} {2022})}\BibitemShut {NoStop}%
\bibitem [{\citenamefont {Jiang}\ \emph {et~al.}(2022)\citenamefont {Jiang},
  \citenamefont {Milo\ifmmode \check{s}\else
  \v{s}\fi{}evi\ifmmode~\acute{c}\else \'{c}\fi{}}, \citenamefont {Wang},
  \citenamefont {Xiao}, \citenamefont {Peeters},\ and\ \citenamefont
  {Chen}}]{Chen2022}%
  \BibitemOpen
  \bibfield  {author} {\bibinfo {author} {\bibfnamefont {J.}~\bibnamefont
  {Jiang}}, \bibinfo {author} {\bibfnamefont {M.}~\bibnamefont {Milo\ifmmode
  \check{s}\else \v{s}\fi{}evi\ifmmode~\acute{c}\else \'{c}\fi{}}}, \bibinfo
  {author} {\bibfnamefont {Y.-L.}\ \bibnamefont {Wang}}, \bibinfo {author}
  {\bibfnamefont {Z.-L.}\ \bibnamefont {Xiao}}, \bibinfo {author}
  {\bibfnamefont {F.}~\bibnamefont {Peeters}}, \ and\ \bibinfo {author}
  {\bibfnamefont {Q.-H.}\ \bibnamefont {Chen}},\ }\href {\doibase
  10.1103/PhysRevApplied.18.034064} {\bibfield  {journal} {\bibinfo  {journal}
  {Phys. Rev. Applied}\ }\textbf {\bibinfo {volume} {18}},\ \bibinfo {pages}
  {034064} (\bibinfo {year} {2022})}\BibitemShut {NoStop}%
\bibitem [{\citenamefont {Daido}\ and\ \citenamefont
  {Yanase}(2022)}]{Daido2022}%
  \BibitemOpen
  \bibfield  {author} {\bibinfo {author} {\bibfnamefont {A.}~\bibnamefont
  {Daido}}\ and\ \bibinfo {author} {\bibfnamefont {Y.}~\bibnamefont {Yanase}},\
  }\href {\doibase 10.48550/ARXIV.2209.03515} {\enquote {\bibinfo {title}
  {Superconducting diode effect and nonreciprocal transition lines},}\ }
  (\bibinfo {year} {2022})\BibitemShut {NoStop}%
\bibitem [{\citenamefont {Kokkeler}\ \emph {et~al.}(2022)\citenamefont
  {Kokkeler}, \citenamefont {Bergeret},\ and\ \citenamefont
  {Golubov}}]{Kokkeler2022}%
  \BibitemOpen
  \bibfield  {author} {\bibinfo {author} {\bibfnamefont {T.}~\bibnamefont
  {Kokkeler}}, \bibinfo {author} {\bibfnamefont {F.~S.}\ \bibnamefont
  {Bergeret}}, \ and\ \bibinfo {author} {\bibfnamefont {A.}~\bibnamefont
  {Golubov}},\ }\href {\doibase 10.48550/ARXIV.2209.13987} {\enquote {\bibinfo
  {title} {Field-free anomalous junction and superconducting diode effect in
  spin split superconductor/topological insulator junctions},}\ } (\bibinfo
  {year} {2022})\BibitemShut {NoStop}%
\bibitem [{\citenamefont {Misaki}\ and\ \citenamefont
  {Nagaosa}(2021)}]{Misaki}%
  \BibitemOpen
  \bibfield  {author} {\bibinfo {author} {\bibfnamefont {K.}~\bibnamefont
  {Misaki}}\ and\ \bibinfo {author} {\bibfnamefont {N.}~\bibnamefont
  {Nagaosa}},\ }\href {\doibase 10.1103/PhysRevB.103.245302} {\bibfield
  {journal} {\bibinfo  {journal} {Phys. Rev. B}\ }\textbf {\bibinfo {volume}
  {103}},\ \bibinfo {pages} {245302} (\bibinfo {year} {2021})}\BibitemShut
  {NoStop}%
\bibitem [{\citenamefont {Hu}(1994)}]{Hu94}%
  \BibitemOpen
  \bibfield  {author} {\bibinfo {author} {\bibfnamefont {C.-R.}\ \bibnamefont
  {Hu}},\ }\href {\doibase 10.1103/PhysRevLett.72.1526} {\bibfield  {journal}
  {\bibinfo  {journal} {Phys. Rev. Lett.}\ }\textbf {\bibinfo {volume} {72}},\
  \bibinfo {pages} {1526} (\bibinfo {year} {1994})}\BibitemShut {NoStop}%
\bibitem [{\citenamefont {Tanaka}\ and\ \citenamefont
  {Kashiwaya}(1995)}]{TK95}%
  \BibitemOpen
  \bibfield  {author} {\bibinfo {author} {\bibfnamefont {Y.}~\bibnamefont
  {Tanaka}}\ and\ \bibinfo {author} {\bibfnamefont {S.}~\bibnamefont
  {Kashiwaya}},\ }\href {\doibase 10.1103/PhysRevLett.74.3451} {\bibfield
  {journal} {\bibinfo  {journal} {Phys. Rev. Lett.}\ }\textbf {\bibinfo
  {volume} {74}},\ \bibinfo {pages} {3451} (\bibinfo {year}
  {1995})}\BibitemShut {NoStop}%
\bibitem [{\citenamefont {Kashiwaya}\ and\ \citenamefont
  {Tanaka}(2000)}]{kashiwaya00}%
  \BibitemOpen
  \bibfield  {author} {\bibinfo {author} {\bibfnamefont {S.}~\bibnamefont
  {Kashiwaya}}\ and\ \bibinfo {author} {\bibfnamefont {Y.}~\bibnamefont
  {Tanaka}},\ }\href {\doibase 10.1088/0034-4885/63/10/202} {\bibfield
  {journal} {\bibinfo  {journal} {Rep. Prog. Phys.}\ }\textbf {\bibinfo
  {volume} {63}},\ \bibinfo {pages} {1641} (\bibinfo {year}
  {2000})}\BibitemShut {NoStop}%
\bibitem [{\citenamefont {L\"ofwander}\ \emph {et~al.}(2001)\citenamefont
  {L\"ofwander}, \citenamefont {Shumeiko},\ and\ \citenamefont
  {Wendin}}]{ABSR2}%
  \BibitemOpen
  \bibfield  {author} {\bibinfo {author} {\bibfnamefont {T.}~\bibnamefont
  {L\"ofwander}}, \bibinfo {author} {\bibfnamefont {V.~S.}\ \bibnamefont
  {Shumeiko}}, \ and\ \bibinfo {author} {\bibfnamefont {G.}~\bibnamefont
  {Wendin}},\ }\href {\doibase 10.1088/0953-2048/14/5/201} {\bibfield
  {journal} {\bibinfo  {journal} {Supercond. Sci. Technol.}\ }\textbf {\bibinfo
  {volume} {14}},\ \bibinfo {pages} {R53} (\bibinfo {year} {2001})}\BibitemShut
  {NoStop}%
\bibitem [{\citenamefont {Yip}(1993)}]{Yip1993}%
  \BibitemOpen
  \bibfield  {author} {\bibinfo {author} {\bibfnamefont {S.}~\bibnamefont
  {Yip}},\ }\href {\doibase 10.1007/BF00120849} {\bibfield  {journal} {\bibinfo
   {journal} {J. Low Temp. Phys.}\ }\textbf {\bibinfo {volume} {91}},\ \bibinfo
  {pages} {203} (\bibinfo {year} {1993})}\BibitemShut {NoStop}%
\bibitem [{\citenamefont {Tanaka}\ and\ \citenamefont
  {Kashiwaya}(1996{\natexlab{a}})}]{TKJosephson}%
  \BibitemOpen
  \bibfield  {author} {\bibinfo {author} {\bibfnamefont {Y.}~\bibnamefont
  {Tanaka}}\ and\ \bibinfo {author} {\bibfnamefont {S.}~\bibnamefont
  {Kashiwaya}},\ }\href {\doibase 10.1103/PhysRevB.53.R11957} {\bibfield
  {journal} {\bibinfo  {journal} {Phys. Rev. B}\ }\textbf {\bibinfo {volume}
  {53}},\ \bibinfo {pages} {R11957} (\bibinfo {year}
  {1996}{\natexlab{a}})}\BibitemShut {NoStop}%
\bibitem [{\citenamefont {Tanaka}\ and\ \citenamefont
  {Kashiwaya}(1997)}]{TKJosephson2}%
  \BibitemOpen
  \bibfield  {author} {\bibinfo {author} {\bibfnamefont {Y.}~\bibnamefont
  {Tanaka}}\ and\ \bibinfo {author} {\bibfnamefont {S.}~\bibnamefont
  {Kashiwaya}},\ }\href {\doibase 10.1103/PhysRevB.56.892} {\bibfield
  {journal} {\bibinfo  {journal} {Phys. Rev. B}\ }\textbf {\bibinfo {volume}
  {56}},\ \bibinfo {pages} {892} (\bibinfo {year} {1997})}\BibitemShut
  {NoStop}%
\bibitem [{\citenamefont {Barash}\ \emph {et~al.}(1996)\citenamefont {Barash},
  \citenamefont {Burkhardt},\ and\ \citenamefont {Rainer}}]{YBJosephson}%
  \BibitemOpen
  \bibfield  {author} {\bibinfo {author} {\bibfnamefont {Y.~S.}\ \bibnamefont
  {Barash}}, \bibinfo {author} {\bibfnamefont {H.}~\bibnamefont {Burkhardt}}, \
  and\ \bibinfo {author} {\bibfnamefont {D.}~\bibnamefont {Rainer}},\ }\href
  {\doibase 10.1103/PhysRevLett.77.4070} {\bibfield  {journal} {\bibinfo
  {journal} {Phys. Rev. Lett.}\ }\textbf {\bibinfo {volume} {77}},\ \bibinfo
  {pages} {4070} (\bibinfo {year} {1996})}\BibitemShut {NoStop}%
\bibitem [{\citenamefont {Testa}\ \emph {et~al.}(2005)\citenamefont {Testa},
  \citenamefont {Sarnelli}, \citenamefont {Monaco}, \citenamefont {Esposito},
  \citenamefont {Ejrnaes}, \citenamefont {Kang}, \citenamefont {Mennema},
  \citenamefont {Tarte},\ and\ \citenamefont {Blamire}}]{Testa}%
  \BibitemOpen
  \bibfield  {author} {\bibinfo {author} {\bibfnamefont {G.}~\bibnamefont
  {Testa}}, \bibinfo {author} {\bibfnamefont {E.}~\bibnamefont {Sarnelli}},
  \bibinfo {author} {\bibfnamefont {A.}~\bibnamefont {Monaco}}, \bibinfo
  {author} {\bibfnamefont {E.}~\bibnamefont {Esposito}}, \bibinfo {author}
  {\bibfnamefont {M.}~\bibnamefont {Ejrnaes}}, \bibinfo {author} {\bibfnamefont
  {D.-J.}\ \bibnamefont {Kang}}, \bibinfo {author} {\bibfnamefont {S.~H.}\
  \bibnamefont {Mennema}}, \bibinfo {author} {\bibfnamefont {E.~J.}\
  \bibnamefont {Tarte}}, \ and\ \bibinfo {author} {\bibfnamefont {M.~G.}\
  \bibnamefont {Blamire}},\ }\href {\doibase 10.1103/PhysRevB.71.134520}
  {\bibfield  {journal} {\bibinfo  {journal} {Phys. Rev. B}\ }\textbf {\bibinfo
  {volume} {71}},\ \bibinfo {pages} {134520} (\bibinfo {year}
  {2005})}\BibitemShut {NoStop}%
\bibitem [{\citenamefont {Tanaka}\ \emph {et~al.}(2009)\citenamefont {Tanaka},
  \citenamefont {Yokoyama},\ and\ \citenamefont {Nagaosa}}]{TYN09}%
  \BibitemOpen
  \bibfield  {author} {\bibinfo {author} {\bibfnamefont {Y.}~\bibnamefont
  {Tanaka}}, \bibinfo {author} {\bibfnamefont {T.}~\bibnamefont {Yokoyama}}, \
  and\ \bibinfo {author} {\bibfnamefont {N.}~\bibnamefont {Nagaosa}},\ }\href
  {\doibase 10.1103/PhysRevLett.103.107002} {\bibfield  {journal} {\bibinfo
  {journal} {Phys. Rev. Lett.}\ }\textbf {\bibinfo {volume} {103}},\ \bibinfo
  {pages} {107002} (\bibinfo {year} {2009})}\BibitemShut {NoStop}%
\bibitem [{\citenamefont {Linder}\ \emph
  {et~al.}(2010{\natexlab{a}})\citenamefont {Linder}, \citenamefont {Tanaka},
  \citenamefont {Yokoyama}, \citenamefont {Sudb\o{}},\ and\ \citenamefont
  {Nagaosa}}]{Linder10a}%
  \BibitemOpen
  \bibfield  {author} {\bibinfo {author} {\bibfnamefont {J.}~\bibnamefont
  {Linder}}, \bibinfo {author} {\bibfnamefont {Y.}~\bibnamefont {Tanaka}},
  \bibinfo {author} {\bibfnamefont {T.}~\bibnamefont {Yokoyama}}, \bibinfo
  {author} {\bibfnamefont {A.}~\bibnamefont {Sudb\o{}}}, \ and\ \bibinfo
  {author} {\bibfnamefont {N.}~\bibnamefont {Nagaosa}},\ }\href {\doibase
  10.1103/PhysRevLett.104.067001} {\bibfield  {journal} {\bibinfo  {journal}
  {Phys. Rev. Lett.}\ }\textbf {\bibinfo {volume} {104}},\ \bibinfo {pages}
  {067001} (\bibinfo {year} {2010}{\natexlab{a}})}\BibitemShut {NoStop}%
\bibitem [{\citenamefont {Lu}\ \emph {et~al.}(2015)\citenamefont {Lu},
  \citenamefont {Yada}, \citenamefont {Golubov},\ and\ \citenamefont
  {Tanaka}}]{LuBo2015}%
  \BibitemOpen
  \bibfield  {author} {\bibinfo {author} {\bibfnamefont {B.}~\bibnamefont
  {Lu}}, \bibinfo {author} {\bibfnamefont {K.}~\bibnamefont {Yada}}, \bibinfo
  {author} {\bibfnamefont {A.~A.}\ \bibnamefont {Golubov}}, \ and\ \bibinfo
  {author} {\bibfnamefont {Y.}~\bibnamefont {Tanaka}},\ }\href {\doibase
  10.1103/PhysRevB.92.100503} {\bibfield  {journal} {\bibinfo  {journal} {Phys.
  Rev. B}\ }\textbf {\bibinfo {volume} {92}},\ \bibinfo {pages} {100503}
  (\bibinfo {year} {2015})}\BibitemShut {NoStop}%
\bibitem [{\citenamefont {Linder}\ \emph
  {et~al.}(2010{\natexlab{b}})\citenamefont {Linder}, \citenamefont {Tanaka},
  \citenamefont {Yokoyama}, \citenamefont {Sudb\o{}},\ and\ \citenamefont
  {Nagaosa}}]{LinderPRB2010}%
  \BibitemOpen
  \bibfield  {author} {\bibinfo {author} {\bibfnamefont {J.}~\bibnamefont
  {Linder}}, \bibinfo {author} {\bibfnamefont {Y.}~\bibnamefont {Tanaka}},
  \bibinfo {author} {\bibfnamefont {T.}~\bibnamefont {Yokoyama}}, \bibinfo
  {author} {\bibfnamefont {A.}~\bibnamefont {Sudb\o{}}}, \ and\ \bibinfo
  {author} {\bibfnamefont {N.}~\bibnamefont {Nagaosa}},\ }\href {\doibase
  10.1103/PhysRevB.81.184525} {\bibfield  {journal} {\bibinfo  {journal} {Phys.
  Rev. B}\ }\textbf {\bibinfo {volume} {81}},\ \bibinfo {pages} {184525}
  (\bibinfo {year} {2010}{\natexlab{b}})}\BibitemShut {NoStop}%
\bibitem [{\citenamefont {Fu}\ and\ \citenamefont {Kane}(2008)}]{FK08}%
  \BibitemOpen
  \bibfield  {author} {\bibinfo {author} {\bibfnamefont {L.}~\bibnamefont
  {Fu}}\ and\ \bibinfo {author} {\bibfnamefont {C.~L.}\ \bibnamefont {Kane}},\
  }\href {\doibase 10.1103/PhysRevLett.100.096407} {\bibfield  {journal}
  {\bibinfo  {journal} {Phys. Rev. Lett.}\ }\textbf {\bibinfo {volume} {100}},\
  \bibinfo {pages} {096407} (\bibinfo {year} {2008})}\BibitemShut {NoStop}%
\bibitem [{\citenamefont {Furusaki}\ and\ \citenamefont
  {Tsukada}(1991{\natexlab{a}})}]{Furusaki91}%
  \BibitemOpen
  \bibfield  {author} {\bibinfo {author} {\bibfnamefont {A.}~\bibnamefont
  {Furusaki}}\ and\ \bibinfo {author} {\bibfnamefont {M.}~\bibnamefont
  {Tsukada}},\ }\href {\doibase https://doi.org/10.1016/0038-1098(91)90201-6}
  {\bibfield  {journal} {\bibinfo  {journal} {Solid State Commun.}\ }\textbf
  {\bibinfo {volume} {78}},\ \bibinfo {pages} {299} (\bibinfo {year}
  {1991}{\natexlab{a}})}\BibitemShut {NoStop}%
\bibitem [{\citenamefont {Bo}\ and\ \citenamefont {Yukio}(2018)}]{LuBo2018}%
  \BibitemOpen
  \bibfield  {author} {\bibinfo {author} {\bibfnamefont {L.}~\bibnamefont
  {Bo}}\ and\ \bibinfo {author} {\bibfnamefont {T.}~\bibnamefont {Yukio}},\
  }\href {http://doi.org/10.1098/rsta.2015.0246} {\bibfield  {journal}
  {\bibinfo  {journal} {Phil. Trans. R. Soc. A.}\ }\textbf {\bibinfo {volume}
  {376}},\ \bibinfo {pages} {20150246} (\bibinfo {year} {2018})}\BibitemShut
  {NoStop}%
\bibitem [{\citenamefont {Tanaka}\ and\ \citenamefont
  {Kashiwaya}(2000)}]{Tanaka2000}%
  \BibitemOpen
  \bibfield  {author} {\bibinfo {author} {\bibfnamefont {Y.}~\bibnamefont
  {Tanaka}}\ and\ \bibinfo {author} {\bibfnamefont {S.}~\bibnamefont
  {Kashiwaya}},\ }\href {\doibase 10.1143/JPSJ.69.1152} {\bibfield  {journal}
  {\bibinfo  {journal} {J. Phys. Soc. Jpn.}\ }\textbf {\bibinfo {volume}
  {69}},\ \bibinfo {pages} {1152} (\bibinfo {year} {2000})}\BibitemShut
  {NoStop}%
\bibitem [{\citenamefont {Furusaki}\ and\ \citenamefont
  {Tsukada}(1991{\natexlab{b}})}]{FT1991b}%
  \BibitemOpen
  \bibfield  {author} {\bibinfo {author} {\bibfnamefont {A.}~\bibnamefont
  {Furusaki}}\ and\ \bibinfo {author} {\bibfnamefont {M.}~\bibnamefont
  {Tsukada}},\ }\href {\doibase 10.1103/PhysRevB.43.10164} {\bibfield
  {journal} {\bibinfo  {journal} {Phys. Rev. B}\ }\textbf {\bibinfo {volume}
  {43}},\ \bibinfo {pages} {10164} (\bibinfo {year}
  {1991}{\natexlab{b}})}\BibitemShut {NoStop}%
\bibitem [{\citenamefont {Beenakker}\ and\ \citenamefont {van
  Houten}(1991)}]{Beenakker91}%
  \BibitemOpen
  \bibfield  {author} {\bibinfo {author} {\bibfnamefont {C.~W.~J.}\
  \bibnamefont {Beenakker}}\ and\ \bibinfo {author} {\bibfnamefont
  {H.}~\bibnamefont {van Houten}},\ }\href {\doibase
  10.1103/PhysRevLett.66.3056} {\bibfield  {journal} {\bibinfo  {journal}
  {Phys. Rev. Lett.}\ }\textbf {\bibinfo {volume} {66}},\ \bibinfo {pages}
  {3056} (\bibinfo {year} {1991})}\BibitemShut {NoStop}%
\bibitem [{\citenamefont {{H. J. Kwon}}\ \emph {et~al.}(2004)\citenamefont {{H.
  J. Kwon}}, \citenamefont {{K. Sengupta}},\ and\ \citenamefont {{V. M.
  Yakovenko}}}]{Yakovenko}%
  \BibitemOpen
  \bibfield  {author} {\bibinfo {author} {\bibnamefont {{H. J. Kwon}}},
  \bibinfo {author} {\bibnamefont {{K. Sengupta}}}, \ and\ \bibinfo {author}
  {\bibnamefont {{V. M. Yakovenko}}},\ }\href {\doibase
  10.1140/epjb/e2004-00066-4} {\bibfield  {journal} {\bibinfo  {journal} {Eur.
  Phys. J. B}\ }\textbf {\bibinfo {volume} {37}},\ \bibinfo {pages} {349}
  (\bibinfo {year} {2004})}\BibitemShut {NoStop}%
\bibitem [{\citenamefont {Tanaka}\ and\ \citenamefont
  {Kashiwaya}(1996{\natexlab{b}})}]{TK96}%
  \BibitemOpen
  \bibfield  {author} {\bibinfo {author} {\bibfnamefont {Y.}~\bibnamefont
  {Tanaka}}\ and\ \bibinfo {author} {\bibfnamefont {S.}~\bibnamefont
  {Kashiwaya}},\ }\href {\doibase 10.1103/PhysRevB.53.9371} {\bibfield
  {journal} {\bibinfo  {journal} {Phys. Rev. B}\ }\textbf {\bibinfo {volume}
  {53}},\ \bibinfo {pages} {9371} (\bibinfo {year}
  {1996}{\natexlab{b}})}\BibitemShut {NoStop}%
\bibitem [{\citenamefont {Akhmerov}\ \emph {et~al.}(2009)\citenamefont
  {Akhmerov}, \citenamefont {Nilsson},\ and\ \citenamefont
  {Beenakker}}]{ANB09}%
  \BibitemOpen
  \bibfield  {author} {\bibinfo {author} {\bibfnamefont {A.~R.}\ \bibnamefont
  {Akhmerov}}, \bibinfo {author} {\bibfnamefont {J.}~\bibnamefont {Nilsson}}, \
  and\ \bibinfo {author} {\bibfnamefont {C.~W.~J.}\ \bibnamefont {Beenakker}},\
  }\href {\doibase 10.1103/PhysRevLett.102.216404} {\bibfield  {journal}
  {\bibinfo  {journal} {Phys. Rev. Lett.}\ }\textbf {\bibinfo {volume} {102}},\
  \bibinfo {pages} {216404} (\bibinfo {year} {2009})}\BibitemShut {NoStop}%
\bibitem [{\citenamefont {Law}\ \emph {et~al.}(2009)\citenamefont {Law},
  \citenamefont {Lee},\ and\ \citenamefont {Ng}}]{LLN09}%
  \BibitemOpen
  \bibfield  {author} {\bibinfo {author} {\bibfnamefont {K.~T.}\ \bibnamefont
  {Law}}, \bibinfo {author} {\bibfnamefont {P.~A.}\ \bibnamefont {Lee}}, \ and\
  \bibinfo {author} {\bibfnamefont {T.~K.}\ \bibnamefont {Ng}},\ }\href
  {\doibase 10.1103/PhysRevLett.103.237001} {\bibfield  {journal} {\bibinfo
  {journal} {Phys. Rev. Lett.}\ }\textbf {\bibinfo {volume} {103}},\ \bibinfo
  {pages} {237001} (\bibinfo {year} {2009})}\BibitemShut {NoStop}%
\bibitem [{\citenamefont {Tsuei}\ \emph {et~al.}(1994)\citenamefont {Tsuei},
  \citenamefont {Kirtley}, \citenamefont {Chi}, \citenamefont {Yu-Jahnes},
  \citenamefont {Gupta}, \citenamefont {Shaw}, \citenamefont {Sun},\ and\
  \citenamefont {Ketchen}}]{Tsuei1994}%
  \BibitemOpen
  \bibfield  {author} {\bibinfo {author} {\bibfnamefont {C.~C.}\ \bibnamefont
  {Tsuei}}, \bibinfo {author} {\bibfnamefont {J.~R.}\ \bibnamefont {Kirtley}},
  \bibinfo {author} {\bibfnamefont {C.~C.}\ \bibnamefont {Chi}}, \bibinfo
  {author} {\bibfnamefont {L.~S.}\ \bibnamefont {Yu-Jahnes}}, \bibinfo {author}
  {\bibfnamefont {A.}~\bibnamefont {Gupta}}, \bibinfo {author} {\bibfnamefont
  {T.}~\bibnamefont {Shaw}}, \bibinfo {author} {\bibfnamefont {J.~Z.}\
  \bibnamefont {Sun}}, \ and\ \bibinfo {author} {\bibfnamefont {M.~B.}\
  \bibnamefont {Ketchen}},\ }\href {\doibase 10.1103/PhysRevLett.73.593}
  {\bibfield  {journal} {\bibinfo  {journal} {Phys. Rev. Lett.}\ }\textbf
  {\bibinfo {volume} {73}},\ \bibinfo {pages} {593} (\bibinfo {year}
  {1994})}\BibitemShut {NoStop}%
\bibitem [{\citenamefont {Tsuei}\ and\ \citenamefont
  {Kirtley}(2000)}]{Tsuei2000}%
  \BibitemOpen
  \bibfield  {author} {\bibinfo {author} {\bibfnamefont {C.~C.}\ \bibnamefont
  {Tsuei}}\ and\ \bibinfo {author} {\bibfnamefont {J.~R.}\ \bibnamefont
  {Kirtley}},\ }\href {\doibase 10.1103/RevModPhys.72.969} {\bibfield
  {journal} {\bibinfo  {journal} {Rev. Mod. Phys.}\ }\textbf {\bibinfo {volume}
  {72}},\ \bibinfo {pages} {969} (\bibinfo {year} {2000})}\BibitemShut
  {NoStop}%
\bibitem [{\citenamefont {Il'ichev}\ \emph {et~al.}(2001)\citenamefont
  {Il'ichev}, \citenamefont {Grajcar}, \citenamefont {Hlubina}, \citenamefont
  {IJsselsteijn}, \citenamefont {Hoenig}, \citenamefont {Meyer}, \citenamefont
  {Golubov}, \citenamefont {Amin}, \citenamefont {Zagoskin}, \citenamefont
  {Omelyanchouk},\ and\ \citenamefont {Kupriyanov}}]{Ilichev}%
  \BibitemOpen
  \bibfield  {author} {\bibinfo {author} {\bibfnamefont {E.}~\bibnamefont
  {Il'ichev}}, \bibinfo {author} {\bibfnamefont {M.}~\bibnamefont {Grajcar}},
  \bibinfo {author} {\bibfnamefont {R.}~\bibnamefont {Hlubina}}, \bibinfo
  {author} {\bibfnamefont {R.~P.~J.}\ \bibnamefont {IJsselsteijn}}, \bibinfo
  {author} {\bibfnamefont {H.~E.}\ \bibnamefont {Hoenig}}, \bibinfo {author}
  {\bibfnamefont {H.-G.}\ \bibnamefont {Meyer}}, \bibinfo {author}
  {\bibfnamefont {A.}~\bibnamefont {Golubov}}, \bibinfo {author} {\bibfnamefont
  {M.~H.~S.}\ \bibnamefont {Amin}}, \bibinfo {author} {\bibfnamefont {A.~M.}\
  \bibnamefont {Zagoskin}}, \bibinfo {author} {\bibfnamefont {A.~N.}\
  \bibnamefont {Omelyanchouk}}, \ and\ \bibinfo {author} {\bibfnamefont
  {M.~Y.}\ \bibnamefont {Kupriyanov}},\ }\href {\doibase
  10.1103/PhysRevLett.86.5369} {\bibfield  {journal} {\bibinfo  {journal}
  {Phys. Rev. Lett.}\ }\textbf {\bibinfo {volume} {86}},\ \bibinfo {pages}
  {5369} (\bibinfo {year} {2001})}\BibitemShut {NoStop}%
\bibitem [{\citenamefont {Veldhorst}\ \emph {et~al.}(2012)\citenamefont
  {Veldhorst}, \citenamefont {Snelder}, \citenamefont {Hoek}, \citenamefont
  {Gang}, \citenamefont {Guduru}, \citenamefont {Wang}, \citenamefont
  {Zeitler}, \citenamefont {van~der Wiel}, \citenamefont {Golubov},
  \citenamefont {Hilgenkamp},\ and\ \citenamefont {Brinkman}}]{veldhorst12}%
  \BibitemOpen
  \bibfield  {author} {\bibinfo {author} {\bibfnamefont {M.}~\bibnamefont
  {Veldhorst}}, \bibinfo {author} {\bibfnamefont {M.}~\bibnamefont {Snelder}},
  \bibinfo {author} {\bibfnamefont {M.}~\bibnamefont {Hoek}}, \bibinfo {author}
  {\bibfnamefont {T.}~\bibnamefont {Gang}}, \bibinfo {author} {\bibfnamefont
  {V.~K.}\ \bibnamefont {Guduru}}, \bibinfo {author} {\bibfnamefont {X.~L.}\
  \bibnamefont {Wang}}, \bibinfo {author} {\bibfnamefont {U.}~\bibnamefont
  {Zeitler}}, \bibinfo {author} {\bibfnamefont {W.~G.}\ \bibnamefont {van~der
  Wiel}}, \bibinfo {author} {\bibfnamefont {A.~A.}\ \bibnamefont {Golubov}},
  \bibinfo {author} {\bibfnamefont {H.}~\bibnamefont {Hilgenkamp}}, \ and\
  \bibinfo {author} {\bibfnamefont {A.}~\bibnamefont {Brinkman}},\ }\href
  {\doibase 10.1038/nmat3255} {\bibfield  {journal} {\bibinfo  {journal} {Nat.
  Mater.}\ }\textbf {\bibinfo {volume} {11}},\ \bibinfo {pages} {417} (\bibinfo
  {year} {2012})}\BibitemShut {NoStop}%
\bibitem [{\citenamefont {Williams}\ \emph {et~al.}(2012)\citenamefont
  {Williams}, \citenamefont {Bestwick}, \citenamefont {Gallagher},
  \citenamefont {Hong}, \citenamefont {Cui}, \citenamefont {Bleich},
  \citenamefont {Analytis}, \citenamefont {Fisher},\ and\ \citenamefont
  {Goldhaber-Gordon}}]{williams12}%
  \BibitemOpen
  \bibfield  {author} {\bibinfo {author} {\bibfnamefont {J.~R.}\ \bibnamefont
  {Williams}}, \bibinfo {author} {\bibfnamefont {A.~J.}\ \bibnamefont
  {Bestwick}}, \bibinfo {author} {\bibfnamefont {P.}~\bibnamefont {Gallagher}},
  \bibinfo {author} {\bibfnamefont {S.~S.}\ \bibnamefont {Hong}}, \bibinfo
  {author} {\bibfnamefont {Y.}~\bibnamefont {Cui}}, \bibinfo {author}
  {\bibfnamefont {A.~S.}\ \bibnamefont {Bleich}}, \bibinfo {author}
  {\bibfnamefont {J.~G.}\ \bibnamefont {Analytis}}, \bibinfo {author}
  {\bibfnamefont {I.~R.}\ \bibnamefont {Fisher}}, \ and\ \bibinfo {author}
  {\bibfnamefont {D.}~\bibnamefont {Goldhaber-Gordon}},\ }\href {\doibase
  10.1103/PhysRevLett.109.056803} {\bibfield  {journal} {\bibinfo  {journal}
  {Phys. Rev. Lett.}\ }\textbf {\bibinfo {volume} {109}},\ \bibinfo {pages}
  {056803} (\bibinfo {year} {2012})}\BibitemShut {NoStop}%
\bibitem [{\citenamefont {Finck}\ \emph {et~al.}(2014)\citenamefont {Finck},
  \citenamefont {Kurter}, \citenamefont {Hor},\ and\ \citenamefont
  {Van~Harlingen}}]{Finck}%
  \BibitemOpen
  \bibfield  {author} {\bibinfo {author} {\bibfnamefont {A.~D.~K.}\
  \bibnamefont {Finck}}, \bibinfo {author} {\bibfnamefont {C.}~\bibnamefont
  {Kurter}}, \bibinfo {author} {\bibfnamefont {Y.~S.}\ \bibnamefont {Hor}}, \
  and\ \bibinfo {author} {\bibfnamefont {D.~J.}\ \bibnamefont
  {Van~Harlingen}},\ }\href {\doibase 10.1103/PhysRevX.4.041022} {\bibfield
  {journal} {\bibinfo  {journal} {Phys. Rev. X}\ }\textbf {\bibinfo {volume}
  {4}},\ \bibinfo {pages} {041022} (\bibinfo {year} {2014})}\BibitemShut
  {NoStop}%
\bibitem [{\citenamefont {Kurter}\ \emph {et~al.}(2015)\citenamefont {Kurter},
  \citenamefont {Finck}, \citenamefont {Hor},\ and\ \citenamefont
  {Van~Harlingen}}]{Kurter2015}%
  \BibitemOpen
  \bibfield  {author} {\bibinfo {author} {\bibfnamefont {C.}~\bibnamefont
  {Kurter}}, \bibinfo {author} {\bibfnamefont {A.~D.~K.}\ \bibnamefont
  {Finck}}, \bibinfo {author} {\bibfnamefont {Y.~S.}\ \bibnamefont {Hor}}, \
  and\ \bibinfo {author} {\bibfnamefont {D.~J.}\ \bibnamefont
  {Van~Harlingen}},\ }\href {\doibase 10.1038/ncomms8130} {\bibfield  {journal}
  {\bibinfo  {journal} {Nat. Commun.}\ }\textbf {\bibinfo {volume} {6}},\
  \bibinfo {pages} {7130} (\bibinfo {year} {2015})}\BibitemShut {NoStop}%
\bibitem [{\citenamefont {Wiedenmann}\ \emph {et~al.}(2016)\citenamefont
  {Wiedenmann}, \citenamefont {Bocquillon}, \citenamefont {Deacon},
  \citenamefont {Hartinger}, \citenamefont {Herrmann}, \citenamefont
  {Klapwijk}, \citenamefont {Maier}, \citenamefont {Ames}, \citenamefont
  {Br{\"u}ne}, \citenamefont {Gould}, \citenamefont {Oiwa}, \citenamefont
  {Ishibashi}, \citenamefont {Tarucha}, \citenamefont {Buhmann},\ and\
  \citenamefont {Molenkamp}}]{Wiedenmann2016}%
  \BibitemOpen
  \bibfield  {author} {\bibinfo {author} {\bibfnamefont {J.}~\bibnamefont
  {Wiedenmann}}, \bibinfo {author} {\bibfnamefont {E.}~\bibnamefont
  {Bocquillon}}, \bibinfo {author} {\bibfnamefont {R.~S.}\ \bibnamefont
  {Deacon}}, \bibinfo {author} {\bibfnamefont {S.}~\bibnamefont {Hartinger}},
  \bibinfo {author} {\bibfnamefont {O.}~\bibnamefont {Herrmann}}, \bibinfo
  {author} {\bibfnamefont {T.~M.}\ \bibnamefont {Klapwijk}}, \bibinfo {author}
  {\bibfnamefont {L.}~\bibnamefont {Maier}}, \bibinfo {author} {\bibfnamefont
  {C.}~\bibnamefont {Ames}}, \bibinfo {author} {\bibfnamefont {C.}~\bibnamefont
  {Br{\"u}ne}}, \bibinfo {author} {\bibfnamefont {C.}~\bibnamefont {Gould}},
  \bibinfo {author} {\bibfnamefont {A.}~\bibnamefont {Oiwa}}, \bibinfo {author}
  {\bibfnamefont {K.}~\bibnamefont {Ishibashi}}, \bibinfo {author}
  {\bibfnamefont {S.}~\bibnamefont {Tarucha}}, \bibinfo {author} {\bibfnamefont
  {H.}~\bibnamefont {Buhmann}}, \ and\ \bibinfo {author} {\bibfnamefont
  {L.~W.}\ \bibnamefont {Molenkamp}},\ }\href {\doibase 10.1038/ncomms10303}
  {\bibfield  {journal} {\bibinfo  {journal} {Nat. Commun.}\ }\textbf {\bibinfo
  {volume} {7}},\ \bibinfo {pages} {10303} (\bibinfo {year}
  {2016})}\BibitemShut {NoStop}%
\bibitem [{\citenamefont {Zareapour}\ \emph {et~al.}(2012)\citenamefont
  {Zareapour}, \citenamefont {Hayat}, \citenamefont {Zhao}, \citenamefont
  {Kreshchuk}, \citenamefont {Jain}, \citenamefont {Kwok}, \citenamefont {Lee},
  \citenamefont {Cheong}, \citenamefont {Xu}, \citenamefont {Yang},
  \citenamefont {Gu}, \citenamefont {Jia}, \citenamefont {Cava},\ and\
  \citenamefont {Burch}}]{Zareapour2012}%
  \BibitemOpen
  \bibfield  {author} {\bibinfo {author} {\bibfnamefont {P.}~\bibnamefont
  {Zareapour}}, \bibinfo {author} {\bibfnamefont {A.}~\bibnamefont {Hayat}},
  \bibinfo {author} {\bibfnamefont {S.~Y.~F.}\ \bibnamefont {Zhao}}, \bibinfo
  {author} {\bibfnamefont {M.}~\bibnamefont {Kreshchuk}}, \bibinfo {author}
  {\bibfnamefont {A.}~\bibnamefont {Jain}}, \bibinfo {author} {\bibfnamefont
  {D.~C.}\ \bibnamefont {Kwok}}, \bibinfo {author} {\bibfnamefont
  {N.}~\bibnamefont {Lee}}, \bibinfo {author} {\bibfnamefont {S.-W.}\
  \bibnamefont {Cheong}}, \bibinfo {author} {\bibfnamefont {Z.}~\bibnamefont
  {Xu}}, \bibinfo {author} {\bibfnamefont {A.}~\bibnamefont {Yang}}, \bibinfo
  {author} {\bibfnamefont {G.~D.}\ \bibnamefont {Gu}}, \bibinfo {author}
  {\bibfnamefont {S.}~\bibnamefont {Jia}}, \bibinfo {author} {\bibfnamefont
  {R.~J.}\ \bibnamefont {Cava}}, \ and\ \bibinfo {author} {\bibfnamefont
  {K.~S.}\ \bibnamefont {Burch}},\ }\href {\doibase 10.1038/ncomms2042}
  {\bibfield  {journal} {\bibinfo  {journal} {Nat. Commun.}\ }\textbf {\bibinfo
  {volume} {3}},\ \bibinfo {pages} {1056} (\bibinfo {year} {2012})}\BibitemShut
  {NoStop}%
\bibitem [{\citenamefont {Tanaka}\ \emph {et~al.}(2012)\citenamefont {Tanaka},
  \citenamefont {Sato},\ and\ \citenamefont {Nagaosa}}]{Tanaka2012}%
  \BibitemOpen
  \bibfield  {author} {\bibinfo {author} {\bibfnamefont {Y.}~\bibnamefont
  {Tanaka}}, \bibinfo {author} {\bibfnamefont {M.}~\bibnamefont {Sato}}, \ and\
  \bibinfo {author} {\bibfnamefont {N.}~\bibnamefont {Nagaosa}},\ }\href
  {\doibase 10.1143/JPSJ.81.011013} {\bibfield  {journal} {\bibinfo  {journal}
  {Journal of the Physical Society of Japan}\ }\textbf {\bibinfo {volume}
  {81}},\ \bibinfo {pages} {011013} (\bibinfo {year} {2012})}\BibitemShut
  {NoStop}%
\end{thebibliography}%
\end{document}